\documentclass[letterpaper]{article} 
\usepackage{aaai2026}  
\usepackage{times}  
\usepackage{helvet}  
\usepackage{courier}  
\usepackage[hyphens]{url}  
\usepackage{graphicx} 
\usepackage{natbib}  
\usepackage{caption} 
\usepackage{algorithm}
\usepackage{algorithmic}

\usepackage{newfloat}
\usepackage{listings}
\DeclareCaptionStyle{ruled}{labelfont=normalfont,labelsep=colon,strut=off} 
\floatstyle{ruled}
\newfloat{listing}{tb}{lst}{}
\floatname{listing}{Listing}
\usepackage{booktabs} 
\usepackage{multirow} 
\usepackage{adjustbox} 
\usepackage{amsmath}
\usepackage[most]{tcolorbox}
\usepackage{float}

\usepackage[english]{babel}
\usepackage[autostyle, english = american]{csquotes}
\MakeOuterQuote{"}

\title{Enhancing and Scaling Search Query Datasets\\for Recommendation Systems}
\author {
    Aaron Rodrigues\textsuperscript{\rm 1},
    Mahmood Hegazy\textsuperscript{\rm 1},
    Azzam Naeem\textsuperscript{\rm 1}
}
\affiliations {
    \textsuperscript{\rm 1}JPMorganChase\\
    aaron.rodrigues@jpmchase.com, mahmood.hegazy@chase.com, azzam.naeem@jpmorgan.com
}

\begin{document}

\maketitle

\begin{abstract}
This paper presents a deployed, production-grade system designed to enhance and scale search query datasets for intent-based recommendation systems in digital banking. In real-world environments, the growing volume and complexity of user intents create substantial challenges for data management, resulting in suboptimal recommendations and delayed product onboarding. To overcome these challenges, our approach shifts the focus from model-centric enhancements to automated, data-centric strategies. The proposed system integrates three core modules: Synthetic Query Generation, Intent Disambiguation, and Intent Gap Analysis. Synthetic Query Generation produces diverse and realistic user queries. Our experiments reveal no statistically significant difference when using synthetic data for Clinc150, while Banking77 and a proprietary dataset show significant differences. We dig into the underlying factors driving these variations, demonstrating that our approach effectively alleviates the cold start problem (i.e. the challenge of recommending new products with limited historical data). Intent Disambiguation refines broad and overlapping intent categories into precise subintents, achieving an F1 score of 0.863 $\pm$ 0.127 against expert reannotations and leading to clearer differentiation and more precise recommendation mapping. Meanwhile, Intent Gap Analysis identifies latent customer needs by extracting novel intents from unlabeled queries; recovery rates reach up to 71\% in controlled evaluations. Deployed in a live banking environment, our system demonstrates significant improvements in recommendation precision, operational agility, and overall data quality, ultimately delivering enhanced user experiences and strategic business benefits. This work underscores the pivotal role of high-quality, scalable data in modern AI-driven applications and advocates a proactive approach to data enhancement as a key driver of value.
\end{abstract}


\section{Introduction}

Recognizing precise customer intents is crucial in digital platforms where search queries are prevalent for assistance or page navigation. This need is particularly acute in financial services, where AI assistants increasingly serve as frontline customer service channels, handling a wide variety of customer intents spanning account management, payments, security concerns, and other banking functions. At the heart of this challenge lies the data: the raw queries that customers type to navigate financial products and services. When harnessed, these queries can dramatically elevate the customer experience by enabling timely, relevant, and precise recommendations. In contrast, outdated query matching techniques often lead to misaligned service suggestions, resulting in customer frustration and lost opportunities.

This paper addresses the business problem of scaling and refining the data that drives recommendation systems in the banking domain. Our work recognizes that while embedding models offer a robust, scalable means of interpreting user intent, their performance is inherently bounded by the quality and granularity of the underlying data. In a competitive environment where products and services evolve and change, traditional data maintenance strategies that rely on static keyword mappings and fixed business rules quickly become outdated. Consequently, the ability to update and expand intent data is paramount to ensuring customer satisfaction, driving conversion rates, and reducing service friction.

\begin{figure}[t]
\centering
\includegraphics[width=0.9\columnwidth]{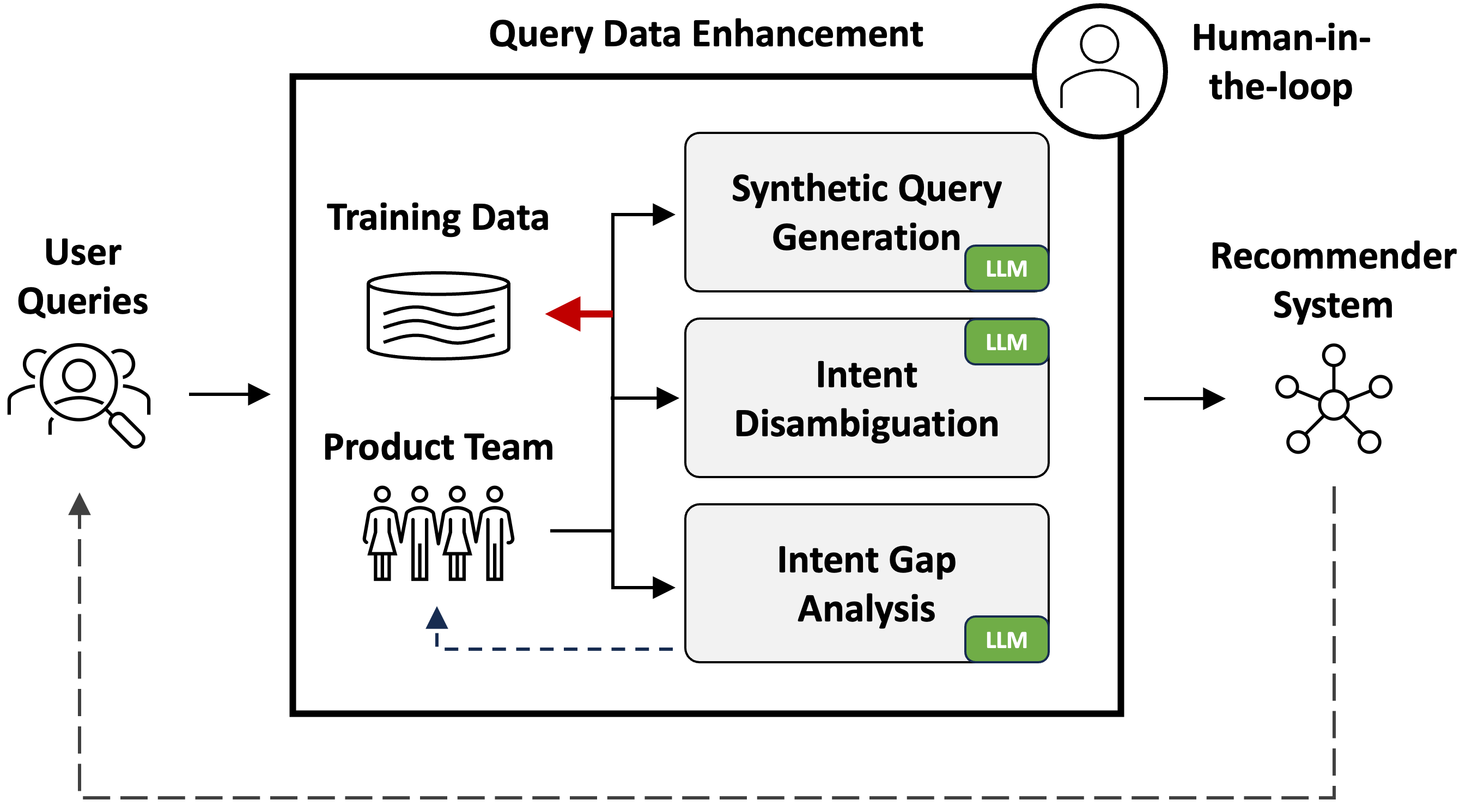}
\caption{Overview of the query data enhancement system. To enhance the user query data used for training intent classifiers, our system employs Synthetic Query Generation and Intent Disambiguation modules. Meanwhile, Intent Gap Analysis identifies intents not captured by the recommender system, to help guide product strategy. The red arrow, pointing back to the dataset, underscores our focus on continuous dataset enrichment to drive performance improvements.}
\label{fig:overview}
\end{figure}

Our proposed system is designed to help manage and enhance the query data feeding embedding-based recommender systems. We propose three complementary modules: Synthetic Query Generation, Intent Disambiguation, and Intent Gap Analysis (Figure \ref{fig:overview}). Each module is tailored to address a specific challenge within the recommendation pipeline. For instance, Synthetic Query Generation enables rapid onboarding of new products and services into the system, significantly alleviating the "cold start" problem that typically plagues recommendation models. Intent Disambiguation refines overly broad or overlapping intent labels, thereby enhancing the precision with which user queries are categorized. Lastly, Intent Gap Analysis serves as a diagnostic tool, revealing latent customer needs or emerging trends that fall outside the current scope of the model.

The ability to dynamically refine and enrich the user query dataset translates directly into business value. Imagine a scenario where a new financial product is introduced (i.e. a cold start). Traditional methods may struggle to incorporate this new product into the recommendation engine until sufficient labeled data accumulates, which can delay deployment and reduce competitive advantage. Our system, on the other hand, integrates synthetic data generation and continuous label refinement, ensuring that new offerings are recommended to customers almost immediately. This speeds up the time-to-market for new products and services.

Moreover, our method provides significant operational benefits. Maintenance of the recommendation system data, which traditionally requires substantial manual intervention, becomes a streamlined, data-driven process. By structuring the user query dataset as a "living document" that is continuously updated and expanded, the need for costly overhauls is minimized. Automated modules backed by Large Language Models (LLMs) ensure that data quality improves over time, even as customer behavior and business objectives evolve. This translates into lower administrative overhead, more efficient deployment cycles, and ultimately, a more resilient recommendation engine.

By shifting the emphasis from model development to data management, we unlock the latent potential within legacy datasets and position recommendation systems to grow with the dynamic digital economy.

\section{Related Work}
To address the cold start problem, we generate synthetic user query data. Previous studies have explored using synthetic text data to supplement real data, and emphasize the capability of LLMs to generate data that closely resembles user interactions in domains where real user data is limited or unavailable. \citet{sahu_data_2022} demonstrated that using LLMs for generating synthetic data in intent classification tasks improves performance when intents are distinct, but semantically close intents led to limitations. \citet{Li_Zhu_Lu_Yin_2023} and \citet{Li_Bonatti_Abdali_Wagle_Koishida_2024} observed that few-shot synthetic data generation improves model performance compared to zero-shot, while noting the impact of data diversity and optimal data volume on classification effectiveness. \citet{dai_auggpt_2025} and \citet{fang_chatgpt_2023} highlighted the use of LLMs for augmenting existing data to improve generalization to different sentence compositions, showing enhanced model performance.

Strategies to evaluate synthetic data generation have been a focus of recent research \citep{chim_evaluating_2025, shaib_detection_2024, tevet_evaluating_2021}. Extrinsic evaluations are considered the gold standard, where classifiers are trained on synthetic data and assessed on a real test set \citep{Li_Zhu_Lu_Yin_2023, Li_Bonatti_Abdali_Wagle_Koishida_2024, fang_chatgpt_2023, chim_evaluating_2025}. This is our primary evaluation method as it aligns with the objective of this study. There is also a growing need for reliable intrinsic evaluations that assess text data quality without the need for model training. Our approach uses a comprehensive set of diversity and complexity measures such as query length, distinct-\emph{N} \citep{li_diversity-promoting_2016}, query mean specificity \citep{hutchison_empirical_2010}, compression ratio \citep{shaib_detection_2024}, and compression ratio - part of speech \citep{shaib_detection_2024}.

To address the problem of new intent discovery, prior work by \citet{Song_He_Wang_Dong_Mou_Wang_Xian_Cai_Xu_2023} evaluated LLM capabilities and proposed methods such as Direct Clustering, Zero Shot Discovery, and Few Shot Discovery. \citet{Rodriguez_Botzer_Vazquez_Pal_Pedersoli_Laradji_2024} used LLMs for intent discovery through automated annotation, highlighting the adaptability of LLMs to diverse user needs. However, given the vast quantities of customer requests received by modern AI assistants, there is a critical need to rank discovered customer intents by query support. This ranking is essential because it directly identifies which missing products or services are most frequently requested by users, providing actionable insights that go beyond intent discovery and enable prioritization based on real demand.

Unlike traditional embedding-based intent discovery methods that rely on well-segmented intents and clear queries \cite{Zhang_Yan_Yang_Ren_Bai_Li_Li_2024, Fan_Pu_Zhang_Wu_2024}, our Intent Gap Analysis module leverages query support and LLM-driven topic modeling to identify and prioritize latent customer needs from noisy, vague queries. LLMs offer enhanced human interpretability and contextual awareness, enabling the generation of a large number of topics that can be refined and merged into coherent and relevant groups \citep{Wang_Prakash_Hoang_Hee_Naseem_Lee_2023, Mu_Dong_Bontcheva_Song_2024, Pham_Hoyle_Sun_Resnik_Iyyer_2024}. Recent work on using LLMs for topic modeling in datasets with short documents have shown promise, particularly since traditional co-occurrence based methods struggle in these domains \citep{Wang_Prakash_Hoang_Hee_Naseem_Lee_2023, Doi_Isonuma_Yanaka_2024, Li_Wang_Zhang_Sun_Ma_2016, Zuo_Wu_Zhang_Lin_Wang_Xu_Xiong_2016, Wu_Li_Zhu_Miao_2020, Qiang_Qian_Li_Yuan_Wu_2022, Laureate_Buntine_Linger_2023, Wu_Luu_Dong_2022}.

\section{Methods}
\subsection{Datasets}

In this study, we evaluated our methods using three intent-labeled query datasets:

\begin{itemize}
    \item \textbf{Clinc150}: Contains 150 intent classes across various domains \citep{Larson_Mahendran_Peper_Clarke_Lee_Hill_Kummerfeld_Leach_Laurenzano_Tang_et_al._2019}. 15k train and 4.5k test queries (balanced).
    
    \item \textbf{Banking77}: Contains 77 intent classes focused on the banking sector \citep{Casanueva_Temčinas_Gerz_Henderson_Vulić_2020}. 10k train and 3.1k test queries (imbalanced).
    
    \item \textbf{Chat146}: A proprietary dataset with 146 intent classes focused on the banking sector. Contains human-annotated queries sampled from financial chatbot conversations and MTurk. 51.3k train and 7.7k test queries (highly imbalanced).
\end{itemize}

Additionally, in deployment analyses we used two more datasets:

\begin{itemize}
    \item \textbf{Chat186}: Contains 56.6k training and 8.6k test queries. Compared to Chat146, it adds 10.3\% more queries (from Synthetic Data Generation), and 6.0k of the training queries were reannotated (from Intent Disambiguation).
    \item \textbf{ChatProd}: Proprietary, unlabeled queries from a financial search service. Filtered to unique queries with at least 3 words that did not lead to any user interaction (i.e., dead ends). 28.5k queries (154k queries before filtering).
\end{itemize}

\begin{figure}[t]
\centering
\includegraphics[width=0.9\columnwidth]{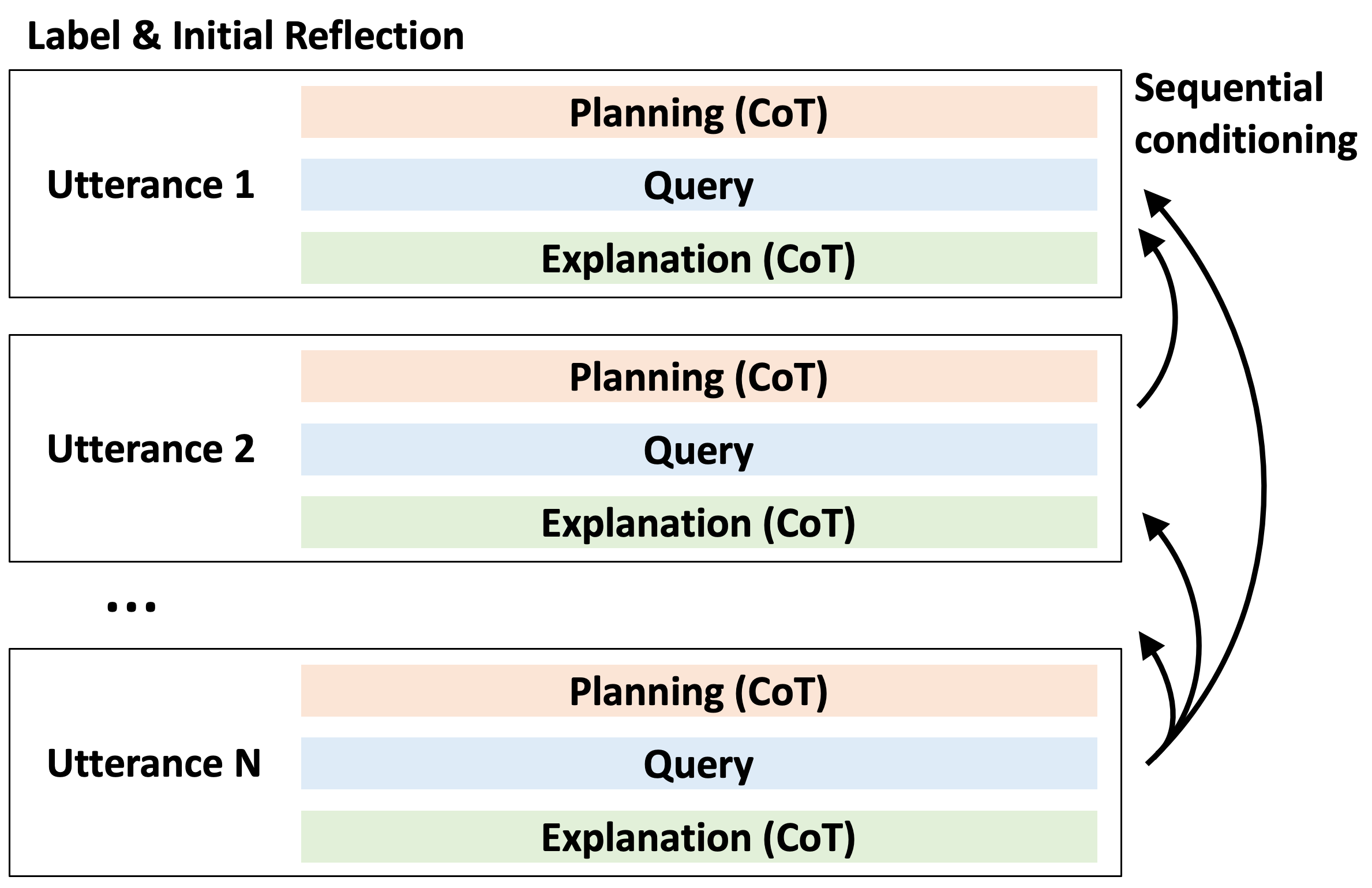}
\caption{Synthetic query generation prompt approach. By conditioning each query to differ from the ones generated earlier, this method produces a spectrum of expressions to capture the variability and diversity of real user queries.}
\label{fig:generation}
\end{figure}

\begin{figure}[t]
\centering
\includegraphics[width=0.95\columnwidth]{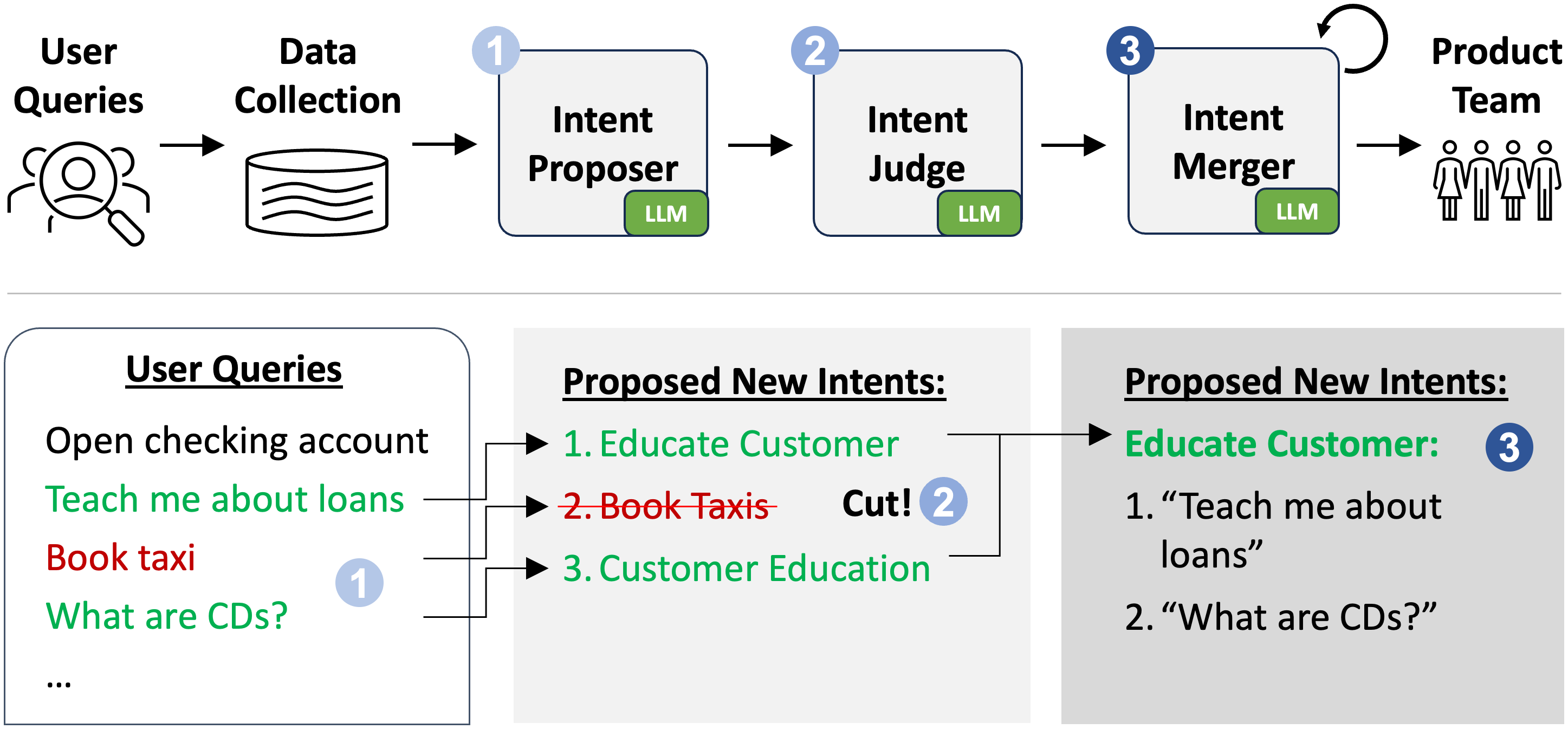}
\caption{Intent Gap Analysis flow and example. Starting with a set of established intents, our process identifies latent customer needs from unlabeled user queries. The Intent Proposer extracts new potential intents. The Intent Judge evaluates each proposal for relevance. The Intent Merger iteratively consolidates similar proposals to highlight the most supported new intents.}
\label{fig:gap_analysis}
\end{figure}

\begin{figure*}[!t]
\centering
\includegraphics[width=0.8\textwidth]{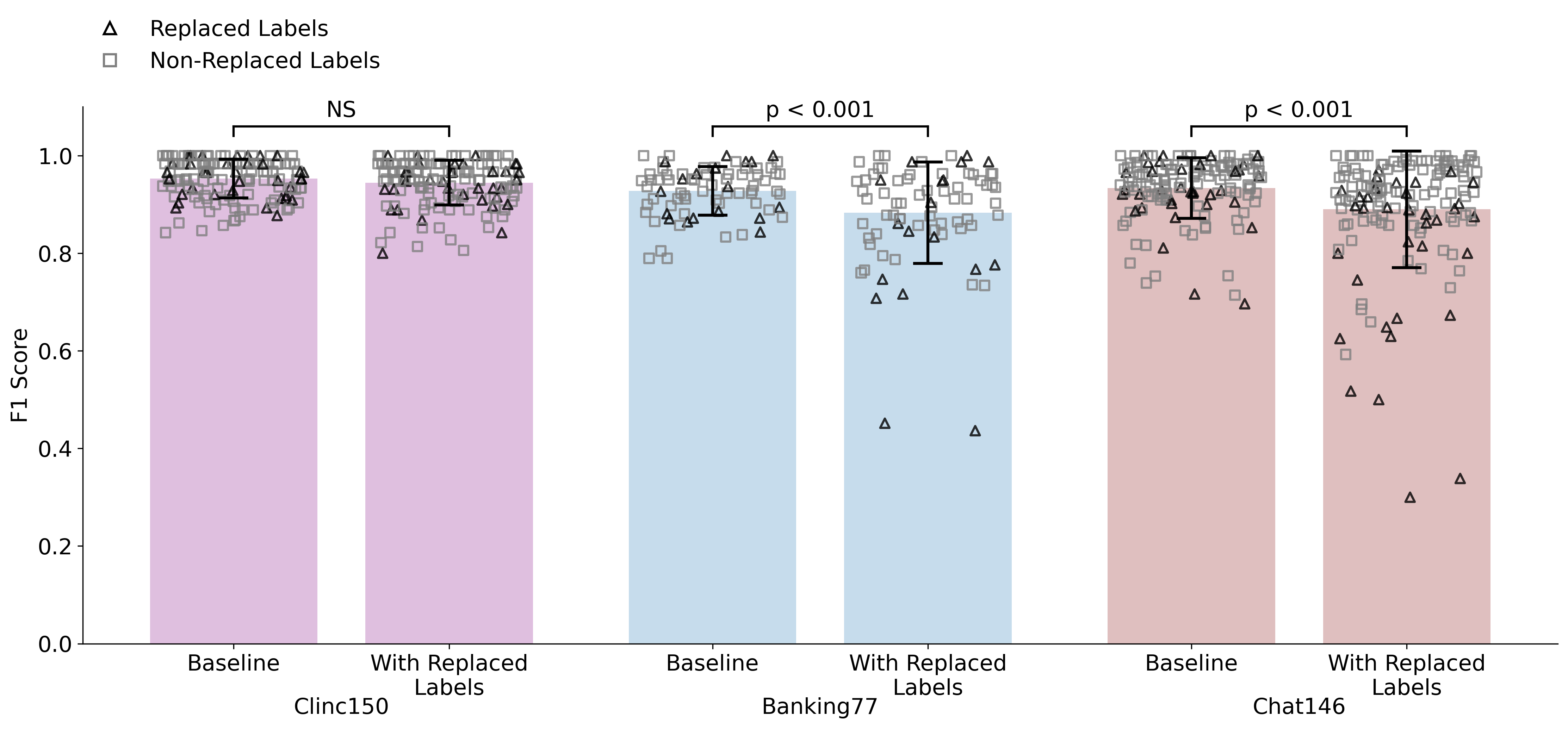}
\caption{Comparison of macro-averaged F1 scores between models trained on control data and data with synthetic data replacement for entire intent labels (mean $\pm$ std). Unchanged classes (gray squares) and fully replaced classes (black triangles) highlight the impact of synthetic data across datasets. Overall, our synthetic queries are comparable to real queries, though they exhibit limitations in specificity, overlap with existing intents, capturing complex intent structures, and linguistic variability.}
\label{fig:generation_results}
\end{figure*}

\subsection{Models}
\label{Models}
The LLM used was GPT4o 2024-08-06 used from June to August 2025. A fixed temperature setting of 0.5 was used; experiments on the effect of temperature are provided in Figure \ref{fig:generation_results_temp}. All prompts are provided in the Appendix.

To train intent classifiers, we fine-tuned DistilBERT models with a classification layer \cite{Sanh_Debut_Chaumond_Wolf_2020}. Class weights were incorporated to address class imbalance. Hyperparameters were fixed and shared across models (learning rate of 1e-4; weight decay of 1e-3; 15 epochs). Performance was evaluated by the macro-averaged F1 score, which treated all classes equally independent of support. 

\subsection{Statistics}
Results are reported as mean ± standard deviation. Per-intent F1 scores were compared using Welch’s two-sample t-test (unequal variances, n = number of intents).

\subsection{Synthetic Query Generation}

We present a novel synthetic query generation method designed to enhance intent classification in banking applications. Our approach generates diverse, realistic user queries by leveraging sequential conditioning (see Figure \ref{fig:generation}).

The generation process begins by selecting few-shot examples: 10 randomly chosen examples from the entire dataset and 10 from the target label. This dual selection not only captures broad query variability but also ensures alignment with the specific intent.

A key innovation is generating multiple queries in a batch, where each subsequent query is explicitly instructed to differ from the previously generated examples (i.e., sequential conditioning). Combined with guidelines to vary the length, complexity, and style, this approach yields a diverse set of queries that reflect the wide range of user expressions for a given intent. In this study, we generate 5 queries per batch and 100 queries per intent in a parallel fashion. The prompt used is provided in Section \ref{synth_prompt}.

\subsubsection{Extrinsic Evaluation}
Extrinsic evaluation was performed using a text classification task to assess the impact of synthetic data on model performance. In our approach, rather than merely augmenting the real dataset with synthetic queries, we mask 25\% of the intent labels in the training set and fully replace the corresponding queries with 100 synthetic queries per intent. This simulates a "cold start" scenario for those intents, testing whether the synthetic data can substitute for real queries. Model effectiveness is measured using macro-averaged F1 scores on a test set comprised solely of real queries. (Traditional data augmentation experiments are provided in Figure \ref{fig:generation_results_rows}.)

DistilBERT models were trained for intent classification (see Section \ref{Models}). By comparing their macro-averaged F1 scores on test sets containing real data, we identified specific intents where synthetic data yields comparable performance, as well as others that may require further refinement. Figure \ref{fig:generation_results} compares models trained on real data with those trained after completely replacing 25\% of intent labels with synthetic data across the three datasets.

Further experiments and ablations on generation batch size, chain-of-thought, temperature, intent descriptions, and few-shot examples are detailed in Figures~\ref{fig:generation_results_batch}--\ref{fig:generation_results_tax}.

\subsubsection{Intrinsic Evaluation}
Our intrinsic evaluation examines the quality and diversity of the synthetic text, which comprises 100 synthetic queries per intent. Query length measures the average length of user queries, reflecting either complexity or conciseness. Distinct-\emph{N} is calculated as the ratio of unique n-grams to the total number of n-grams, providing a quantitative measure of diversity \citep{li_diversity-promoting_2016, li_contrastive_2023}. Self-BLEU quantifies corpus diversity by computing the similarity of each query against all others, with lower scores indicating higher diversity \citep{zhu_texygen_2018}. Query mean specificity (QMS) evaluates the uniqueness of terms in the corpus, with higher values indicating greater specificity \citep{hutchison_empirical_2010}. The compression ratio assesses text complexity by comparing the size of the compressed text to the original, and the compression ratio - part of speech evaluates the diversity of sentence structures by comparing the size of compressed POS tag sequences to the original. Results are presented in Table \ref{tab:intrinsic_synth_table}, and corresponding formulas are detailed in Appendix \ref{intrinsic_sd_eval}.

\subsection{Intent Disambiguation}

Our intent disambiguation approach leverages LLM capabilities to reclassify and refine broad intent labels into more specific subintents using a provided set of examples. For each broad intent, the process employs 10 few-shot examples per subintent to guide the LLM in accurately labeling user queries. A default label is applied for ambiguous cases to ensure that, when the LLM lacks confidence, the original broad category is retained. Additionally, the process is designed to be parallelizable (queries were re-annotated in batches of 25). The disambiguation prompt is provided in Section \ref{id_prompt}.

\subsubsection{Evaluation}

We evaluated the Intent Disambiguation module by comparing 6.0k Chat146 queries reannotated by humans into more precise subintents with the LLM's reannotations. The results are shown in Figure~\ref{fig:reannotation_results}.

\subsection{Intent Gap Analysis}

In this section, we aim to uncover latent customer intents that remain unaddressed by our current taxonomy. To achieve this, we prepared a novel, agentic three-phase process that identifies, evaluates, and ranks new intents based on query support:

\begin{itemize}
    \item \textbf{Propose Intents:} LLMs analyzed batches of 25 queries in comparison to the existing set of intents. LLMs evaluated whether certain queries fall outside the established categories, providing a detailed chain-of-thought explanation along with proposals for new, actionable intents.
    \item \textbf{Judge Proposed Intents:} A second layer of LLM analysis judged each proposed intent against the existing intent set. Each new intent was categorized as "not novel," "not relevant," "FAQ," or "consider adding." This ensured that only intents which are novel, significant, and directly related to user actions were carried forward.
    \item \textbf{Merge Intents:} The validated proposed intents were merged iteratively using LLMs to resolve semantic redundancies. In each iteration, if two proposed intents were found to express the same customer need, they were merged into a single proposal that aggregates their query support. This merging process was performed for 2000 iterations, resulting in a final ranked list where intents are ordered by query support, highlighting emergent customer needs not captured by the current intents.
\end{itemize}

Prompts for each phase are provided in Section \ref{iga_prompts}. Additional experiments on the number of Merge Intent iterations are provided in Figure \ref{fig:rediscovery_results_all}.

\subsubsection{Evaluation}

For evaluation, we reframed the discovery task into a rediscovery problem. Instead of directly assessing whether new intents were discovered, we purposely masked 25\% of the known intents and then applied Intent Gap Analysis. This approach allowed us to assess whether these hidden intents could be recovered compared to a baseline.

In the baseline method, LLMs were provided with the masked taxonomy (i.e., the masked intents along with 10 few-shot examples each) and instructed to generate 10 new customer intents, complete with descriptions and supporting examples.\footnote{We would have generated 100 intents directly but could not due to context window limitations.} The generated intents were then aggregated and formatted to mirror the output structure of the gap analysis.

Both methods used the same evaluation pipeline. For each masked intent, an LLM judge compared its examples with those of the proposed intents to determine if an exact semantic match existed. Each judgement was repeated 5 times to account for variability. From these responses, we computed the \emph{intent recovery rate}: the percentage of masked intents rediscovered among the top N proposals (Figure \ref{fig:gap_analysis}), with N ranging from 10 to 100. 

\section{Deployment}

This system supports the data for training a search recommender system. By addressing underlying data challenges, the deployment of our modules has streamlined problem resolution and boosted operational efficiency. Our system enables rapid model adjustments that are more adaptive than traditional keyword and lexical matching methods.

A key challenge we address is the cold start problem, where the lack of annotated queries for new webpages results in costly human annotation and delays until sufficient real queries become available. Recent shifts from keyword-based matching to intent-driven recommendations have further highlighted the need for a robust solution to close these data gaps. Our Synthetic Data Generation module alleviates this issue by generating diverse and realistic user queries for underrepresented intents.

Intent Disambiguation emerged from the need to refine training data, particularly within the Chat146 dataset, where broad intent categories were impeding recommendation accuracy.  This module reannotated 6.0k intents into more precise subintents. For example, intents associated with numerous products and services, such as \texttt{Manage Settings}, \texttt{Open Account}, and \texttt{Book Travel}, were diluting recommendation relevance. By breaking these broad intents into granular subintents, our training data more accurately reflects distinct user needs. Additionally, the Synthetic Data Generation module contributed by supplementing underrepresented subintents (such as \texttt{Open Money Market Account}) with realistic user queries.

Intent Gap Analysis, which identifies intents with substantial support from real user queries, is used to help support product strategy. Although this is a slow process and has yet to deliver direct results, the authors are confident in its potential. An example application could be to address out-of-domain user queries with FAQs explaining why the discovered products and services are not available.

Designed for simplicity and efficiency, our web application integrates the three core modules—synthetic query generation, intent disambiguation, and intent gap analysis—into dedicated pages. A data health dashboard displays example queries and their support levels for each intent, enabling a human-in-the-loop to generate, reannotate, review, and submit hundreds of queries to the dataset within seconds. In future work, registering new products and services to the intent dataset will be as simple as uploading mobile app screenshots and verifying the generated synthetic queries.

Future enhancements will focus on refining confidence ranking mechanisms to better support human-in-the-loop processes, ensuring that outputs receive thorough inspection before approval. Additionally, we plan to implement additional checks for overlapping queries across intents before incorporating new ones, complementing the Intent Disambiguation module which focuses on refining subintents within a single broad category.
\label{DepIGA}

\begin{figure}[t]
\centering
\includegraphics[width=0.95\columnwidth]{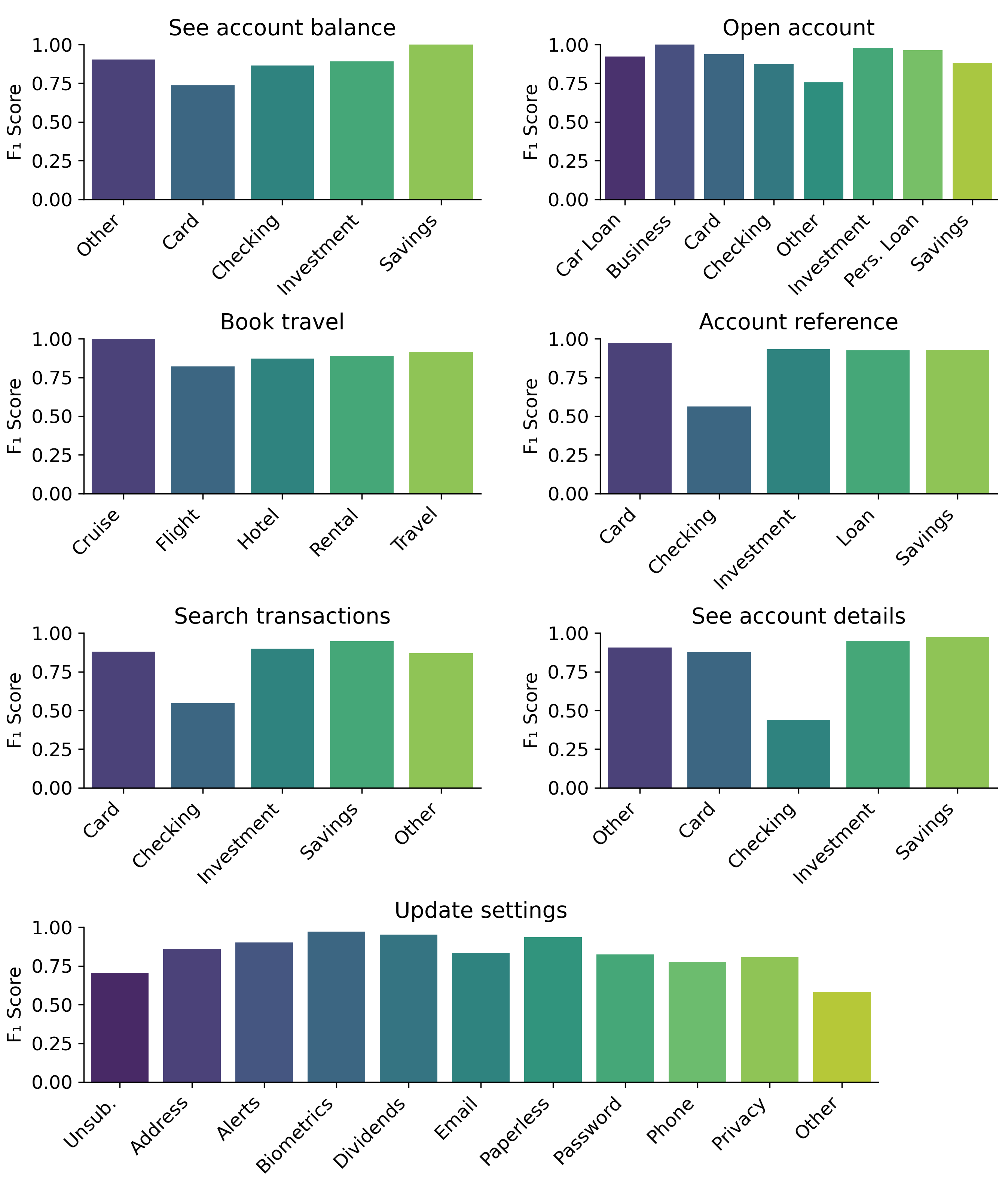}
\caption{Intent disambiguation results (F1 score) comparing an LLM annotator to human reannotations of existing labeled queries into more precise subintents from the Chat146 dataset. Overall, LLM disambiguations aligned well with human judgment, with most discrepancies arising from disagreements over whether "debit card" queries should be classified under Credit Card or Checking subintents.}
\label{fig:reannotation_results}
\end{figure}

\begin{figure}[t]
\centering
\includegraphics[width=0.95\columnwidth]{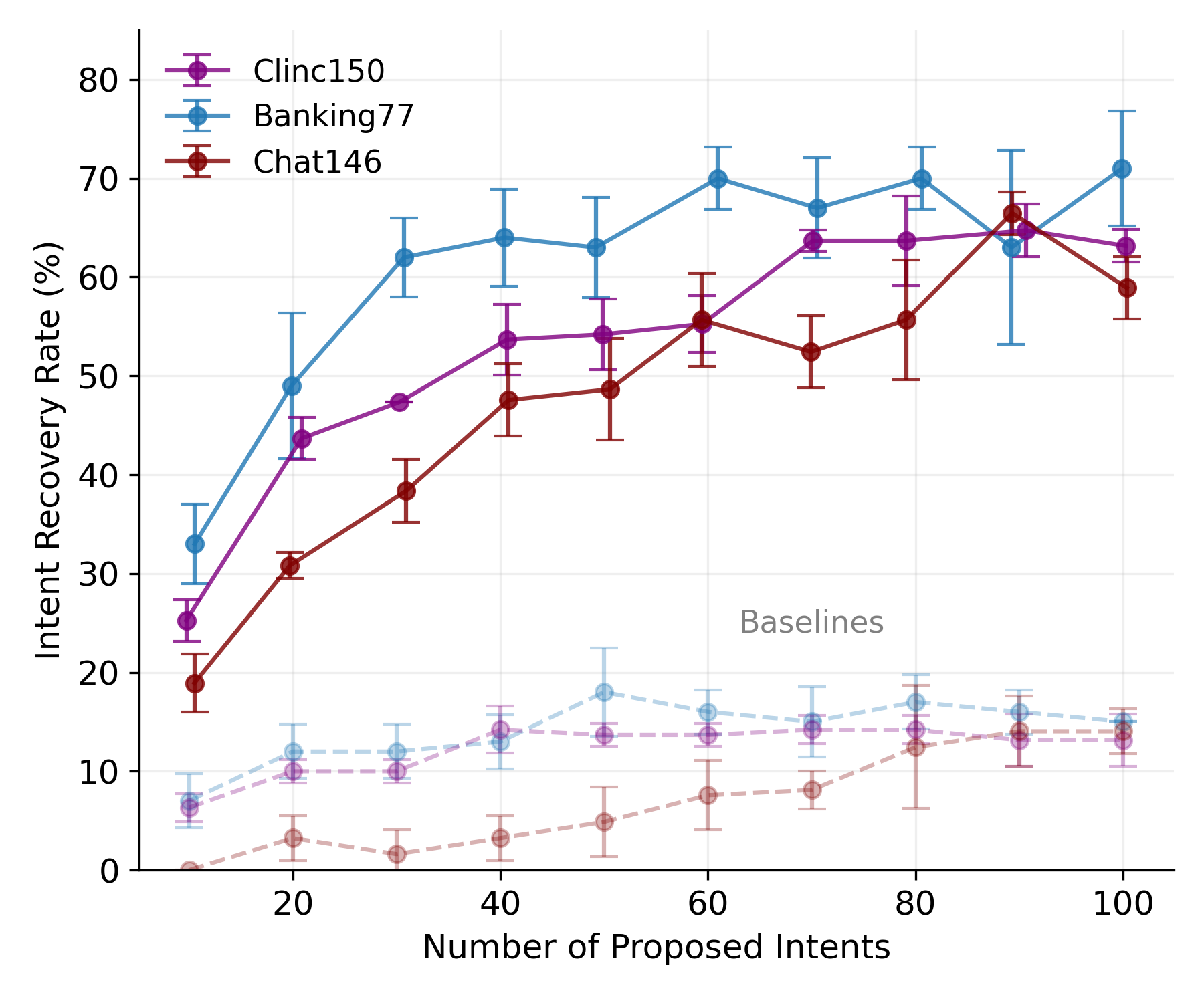}
\caption{Rediscovery rate of masked intents (mean $\pm$ std). After masking 25\% of the intents (with their few-shot examples) from the known set, we applied our method to evaluate how effectively the masked intents can be rediscovered. The x-axis represents the number of proposed intents evaluated for recovery, with proposals ordered by support (i.e. the number of examples per intent). Across three datasets, over 50\% of the masked intents are recovered among the top 60 most supported proposed intents. Beyond 70 proposed intents, the recovery rate plateaus, suggesting that the best proposals are the ones with the most support.}
\label{fig:rediscovery_results}
\end{figure}

\begin{table*}[h]
\centering
\caption{Intrinsic performance metrics between real and synthetic queries. Metrics include Query Length, Distinct-\emph{N}, Self-BLEU, Query Mean Specificity (QMS), Compression Ratio (CR), and Compression Ratio - Part of Speech (CR-POS).}
\label{tab:intrinsic_synth_table}
\begin{adjustbox}{max width=\textwidth}
\begin{tabular}{l c c c c c c}
\toprule
\multirow{2}{*}{\textbf{}} & \multicolumn{2}{c}{\textbf{Clinc150}} & \multicolumn{2}{c}{\textbf{Banking77}} & \multicolumn{2}{c}{\textbf{Chat146}} \\ 
\cmidrule(lr){2-3} \cmidrule(lr){4-5} \cmidrule(lr){6-7}
                                & \textbf{Real} & \textbf{Synthetic} & \textbf{Real} & \textbf{Synthetic} & \textbf{Real} & \textbf{Synthetic} \\
\midrule
\textbf{Query Length}    & 39.9 $\pm$ 15.3  & 28.0 $\pm$ 12.9  & 59.5 $\pm$ 40.9  & 36.1 $\pm$ 18.1  & 44.4 $\pm$ 25.9  & 28.8 $\pm$ 13.1 \\
\textbf{Distinct-\emph{N}}  & 0.354 $\pm$ 0.265 & 0.378 $\pm$ 0.265 & 0.379 $\pm$ 0.304 & 0.385 $\pm$ 0.271 & 0.349 $\pm$ 0.287 & 0.379 $\pm$ 0.265 \\
\textbf{Self-BLEU}      & 0.556 $\pm$ 0.369 & 0.423 $\pm$ 0.428 & 0.593 $\pm$ 0.317 & 0.447 $\pm$ 0.412 & 0.496 $\pm$ 0.384 & 0.371 $\pm$ 0.426 \\
\textbf{QMS}                     & 8.58 $\pm$ 1.40  & 8.48 $\pm$ 1.35  & 7.99 $\pm$ 1.51  & 7.76 $\pm$ 1.37  & 9.96 $\pm$ 1.42  & 8.44 $\pm$ 1.37 \\
\textbf{CR}       & 0.228 & 0.222 & 0.233 & 0.217 & 0.235 & 0.217 \\
\textbf{CR-POS}   & 0.096 & 0.098 & 0.097 & 0.093 & 0.097 & 0.093 \\
\bottomrule
\end{tabular}
\end{adjustbox}
\end{table*}

\section{Results and Discussion}

\subsection{Synthetic Query Generation}


Extrinsic evaluation was performed by simulating a "cold start" scenario, where synthetic queries replaced all real training queries for 25\% of the intent labels. Model performance was measured using macro-averaged F1 scores on a test set of real queries. Figure \ref{fig:generation_results} compares the performance of models trained on control data (with unchanged intent labels, represented by gray squares) against models where the corresponding intent labels were completely replaced by synthetic data (represented by black triangles). In the Banking77 dataset, the baseline trained on real data had a macro-averaged F1 score of 0.928 ± 0.035, whereas the synthetic data model scored 0.883 ± 0.073. For Chat146, the baseline model reached 0.935 ± 0.040 compared to 0.892 ± 0.083 for the synthetic version. In contrast, the Clinc150 dataset showed no significant difference between baseline and synthetic F1 scores, at 0.953 ± 0.028 and 0.944 ± 0.032 respectively. The statistically significant performance decreases observed for Banking77 and Chat146 (p $<$ 0.001) suggest that while synthetic data is effective in mitigating the cold start problem, the method exhibits limitations in more domain-specific and intricate datasets.

The intrinsic evaluation results, shown in Table~\ref{tab:intrinsic_synth_table}, provide insights into the discrepancies between synthetic and real query data. Synthetic queries tend to be shorter and less variable in length. For example, in the Banking77 dataset, real queries average 59.5 $\pm$ 40.9 characters versus 36.1 $\pm$ 18.1 characters for synthetic queries, with similar trends observed in Clinc150 and Chat146. The Distinct-\emph{N} score measures lexical diversity and is higher for synthetic data, indicating a broader range of expressions. Similarly, lower self-BLEU scores for synthetic queries compared to real queries suggest that the synthetic queries exhibit greater internal diversity. The query mean specificity (QMS) was slightly lower in synthetic data when compared to real data, likely because synthetic data lacked specific entities (e.g., names, emails, merchants, account numbers) that users naturally include. The compression ratio (CR) was slightly lower for synthetic data, suggesting that synthetic queries were marginally less complex. Similarly, the CR-POS, which assesses the diversity of sentence structures via part-of-speech sequences, was similar between real and synthetic data, further supporting that the synthetic data preserves structural variety. Despite some differences, the overall intrinsic metrics indicate that synthetic queries maintain a level of diversity, complexity, and structural variation comparable to that of real data.

\begin{figure*}[!t]
\centering
\includegraphics[width=0.80\textwidth]{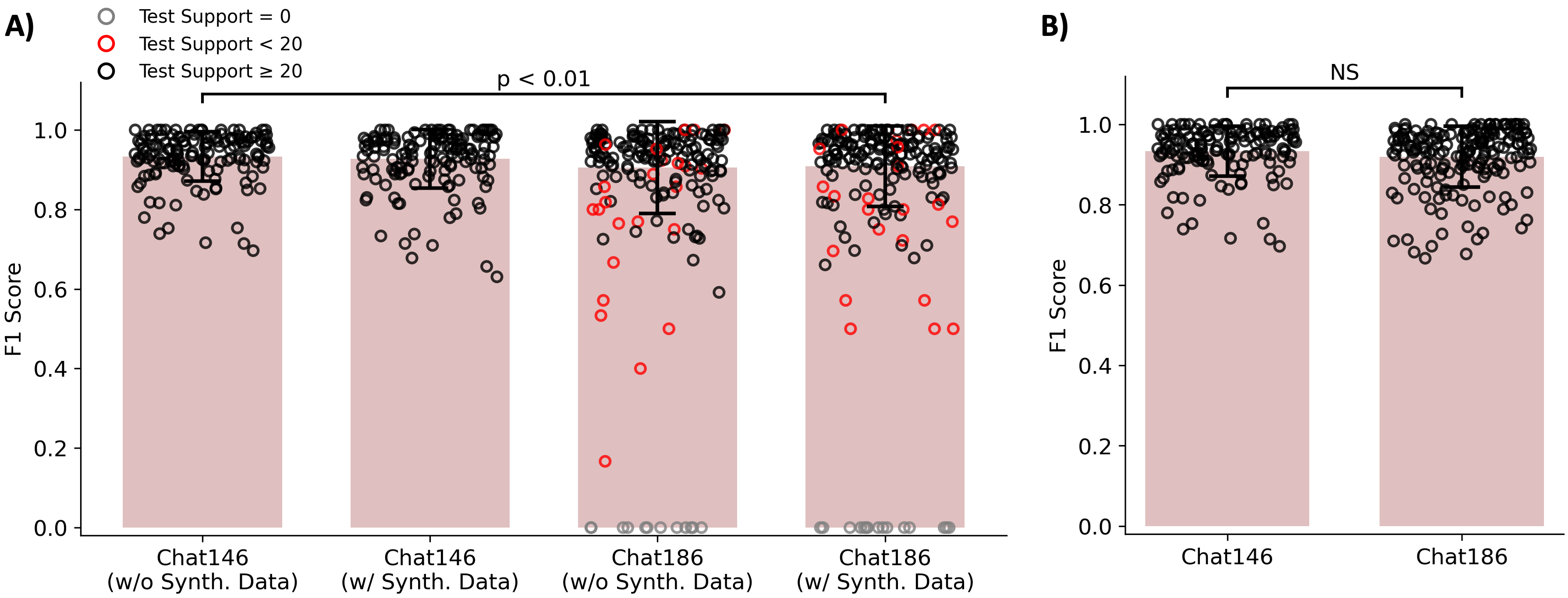}
\caption{Comparison of F1 scores for models trained on Chat146 and Chat186 (mean $\pm$ std). \textbf{Panel A:} Four configurations evaluated on the Chat146 test set: (i) Chat146 with real data only, (ii) Chat146 with synthetic data augmentation, (iii) Chat186 with real data only, and (iv) Chat186 with synthetic data augmentation. Intents with fewer than 20 support queries are outlined in red, while those with zero support (gray) are excluded from the F1 calculation. The results show that incorporating synthetic data after intent disambiguation enhances model performance and highlights gaps in subintent representation within the Chat146 test set. \textbf{Panel B:} F1 scores for models trained on Chat146 and Chat186, where the Chat186 test set includes synthetic queries.}
\label{fig:prod_generation_results}
\end{figure*}

\subsubsection{Deployment}
Chat186 is a production-grade dataset that derives from Chat146 after performing Intent Disambiguation and Synthetic Query Generation. Figure~\ref{fig:prod_generation_results} presents a comparison of the macro-averaged F1 scores achieved by models trained on different datasets and configurations. In Panel A, we evaluate four configurations using the Chat146 test set: (i) Chat146 trained on real data only, (ii) Chat146 augmented with synthetic data, (iii) Chat186 with real data only, and (iv) Chat186. Intents with fewer than 20 support queries in the test set are highlighted in red, whereas those with zero support are in gray and are omitted from the F1 calculation. This analysis demonstrates that incorporating synthetic data after Intent Disambiguation stage can enhance model performance. It also highlights existing gaps in subintent coverage within the Chat146 test set.

Panel B compares the macro-averaged F1 scores between models trained on Chat146 and Chat186, where the Chat186 test set includes synthetic queries. No significant difference was measured between the original and updated models. These findings suggest that Intent Disambiguation and Synthetic Query Generation can be effectively leveraged to expand the number of intents and improve classification precision without adversely affecting model performance.

\subsubsection{Discussion}

Our intrinsic evaluation indicates that while synthetic data closely mirrors many properties of real user queries, some differences remain that affect downstream performance. The slightly higher Distinct-\emph{N} scores in synthetic data may result from LLM-generated typos, which superficially introduce lexical diversity. In terms of query length, synthetic data consistently exhibits shorter and less variable utterances compared to real data. This contrasts with \citet{Sannigrahi_Fraga-Silva_Oualil_Van_Gysel_2024}, where LLM-generated queries were significantly longer than real queries. This is because our prompting strategy emphasizes short queries to mitigate excessive verbosity and ensure compatibility with intents that handle incomplete queries. However, this also means that lengthy queries present in real-world data are underrepresented in our synthetic samples. Notably, the query mean specificity (QMS) is lower in synthetic data across all datasets. This gap likely arises because synthetic queries tend to omit highly specific entities—such as distinctive merchant names, product identifiers, or email addresses—that real users naturally include (e.g., "Warren," "hotmail," "Iberia", "0532"). Instead, LLM-generated queries often feature more common entities (e.g., "Alice," "gmail," "Amazon", "1234"), which may not fully capture the nuances required in domain-specific applications. 

The ability of synthetic data to address cold-start scenarios and the observations from intrinsic analysis are further explored in the extrinsic analysis. In Figure \ref{fig:generation_results}, we observe that representing entire intent classes with synthetic data led to a significant impact on model performance for the Banking77 and Chat146 datasets. Qualitative analysis revealed the following key areas where Synthetic Query Generation fell short:

\begin{itemize}
    \item \textbf{Specificity and Diversity in Queries}: Real queries for intents such as \texttt{New Account Inquiry} were specific and diverse, containing detailed context and varied language. In contrast, synthetic queries were often too generic and sometimes resembled other intents, such as \texttt{Open Account}, which contributed to a drop in F1 scores (e.g., a drop of 0.42 for \texttt{New Account Inquiry} in Chat146).
    
    \item \textbf{Overlap with Existing Intents}: Synthetic queries for intents such as \texttt{Card Lock Status} overlapped with existing intents like \texttt{Lock Card} in Chat146. In Banking77, the intents \texttt{topping\_up\_by\_card} and \texttt{declined\_card\_payment} had the largest F1 score drops of 0.44 and 0.42, respectively, due to overlap with related intents like \texttt{top\_up\_failed} and \texttt{card\_not\_working}. This was similarly observed by \citet{sahu_data_2022}.
    
    \item \textbf{Domain-Spanning Intents with Subintents}: Intents like \texttt{Open Account} and \texttt{Reward Benefit Inquiry} in Chat146 that span multiple products and contain numerous subintents were challenging for the generation process. Synthetic queries failed to capture the full scope and complexity of these intents.
    
    \item \textbf{Human Expressions}: Human queries, especially in the \texttt{Social Talk} and \texttt{Help} intents for Chat146, had a wide range of linguistic variability (e.g. profanity). Synthetic queries, on the other hand, were more uniform.
\end{itemize}

To improve synthetic data generation for complex and nuanced user interactions, future work could integrate several strategies. One approach is to extend the current process with intent-specific prompt tuning and selection of optimal few-shot examples \citep{ye_progen_2022}, which could help generate more context-aware queries. Generating queries with more variable prompts to increase diversity and better capture the full domain of each intent could also be explored \citep{yu_large_2023}.

\subsection{Intent Disambiguation}
Evaluation of the Intent Disambiguation module demonstrates that the LLM-guided annotation approach effectively aligns with human annotations, refining broad intent labels into more precise subintents. 

The single agent annotation approach achieved a macro-averaged F1 score of 0.863 ± 0.127 over 6.0k queries and 44 subintents. Qualitative analysis revealed the following discrepancies between human and LLM annotations:

\begin{enumerate}
    \item \textbf{Category Disagreements}: For example, disagreements arose regarding the classification of "debit cards" queries. LLMs categorized them under the "Card" category (which was intended exclusively for credit cards), while human annotators placed them under "Checking," since debit cards are associated with checking accounts.

    \item \textbf{Queries Spanning Multiple Subintents}: For example, "change my phone and email" should be classified under the "Other" subintent due to its ambiguity, rather than under "Phone" or "Email" subintents.

    \item \textbf{Business Knowledge Requirements}: For example, queries like "Sapphire" and "United" should be classified under "Credit Card" due to its likely reference to a credit card product, but LLMs may lack this business knowledge.
\end{enumerate}

\subsubsection{Deployment}

In preparing the truth set, the authors and an expert linguist spent several hours reannotating queries to refine the initial seven intents into more precise subintents (see Figure \ref{fig:reannotation_results}). Although entity tags and lexical matches were used to accelerate the process, they sometimes fell short, especially for queries with multiple subintents or requiring specific business knowledge (as discussed earlier). Additionally, the sheer volume of data meant that human reviewers could easily overlook annotation errors. For instance, while developing Figure \ref{fig:reannotation_results}, over a hundred cases emerged where the LLM’s annotations were more accurate than the original human annotations. This led to amending the truth set and further reinforced the value of a human-in-the-loop approach, where LLMs perform the initial work before human review.

\begin{table}[h]
\centering
\caption{Snapshot of proposed intents from applying Intent Gap Analysis on ChatProd.}
\label{tab:gap_analysis}
\begin{adjustbox}{max width=\columnwidth}
\begin{tabular}{lll}
\toprule
\textbf{Proposed Intent} & \textbf{Representative Query} & \textbf{Support} \\
\midrule
Request Virtual Card & Make a temp credit card  & 46\\
Manage Payment Methods & Add a new payment method & 36\\
Open Child Account & Open bank account daughter & 30\\
Early Pay & Early payday please & 17\\
Join Account Management & Add joint account holder & 12\\
MFA Setup & Two step sign-in & 11\\
\bottomrule
\end{tabular}
\end{adjustbox}
\end{table}

\subsubsection{Discussion}
The failure modes identified in the evaluation results are addressed in deployment by allowing a human-in-the-loop to provide specific details, nuances, and descriptions about the subintents before running Intent Disambiguation. Additionally, we are working on a Multi-Agent Framework for Annotation approach that is well-suited for this application \cite{Hegazy_Rodrigues_Naeem_2025}, further enhancing the accuracy of the disambiguation process.

\subsection{Intent Gap Analysis}
Figure \ref{fig:rediscovery_results} summarizes the recovery performance across three datasets by comparing the percentage of masked intents that our gap analysis method successfully rediscovered versus a baseline. The x-axis represents the number of proposed intents evaluated for recovery, with proposals ordered by support (i.e. the number of examples per intent). This ordering assumes that higher support indicates greater quality and relevance, which is a hypothesis supported by the results.

For each dataset, the recovery rate increases steeply at lower numbers of proposals and then levels off. For example, in the Banking77 experiments the method recovered 63.0\% $\pm$ 5.1\% of the masked intents when the top 50 proposed intents were considered, and this rate increased to 71.0\% $\pm$ 5.8\% by the top 100 proposals. In the Clinc150 dataset, the recovery rate was 54.2\% $\pm$ 3.6\% using the top 50 proposals, reaching approximately 63.2\% $\pm$ 1.7\% when 100 proposals was considered. The Chat146 dataset showed slightly lower recovery, with 48.7\% $\pm$ 5.1\% recovered at the 50-intent mark and reached 58.9\% $\pm$ 3.2\% when considering 100 proposals.

An analysis of the incremental improvement between successive proposal counts reveals that the most significant gains occur when fewer than 50 proposals are considered. Beyond about 50 proposals for Banking77 and Clinc150, additional intents yield only marginal improvement, signaling a plateau. In contrast, the Chat146 dataset shows marginal improvement only after approximately 70 proposals, likely due to the greater complexity of the intent set.

These findings indicate that our method efficiently rediscovers the majority of masked intents by relying on a relatively small number of high-support proposals. The plateau is likely due to two factors: first, the additional proposals are less supported (and hence lower-quality) intents that do not contribute significantly to further recovery; and second, when a large number of intents (70+) are fed into an LLM as context for judging similarity, context window limitations may affect judgment performance. 

\subsubsection{Deployment}
\label{gap_analysis_prod}

Applying Intent Gap Analysis to the ChatProd dataset has successfully captured latent customer needs and pinpointed areas for new app products and services. For example, the method groups real customer utterances (with high query support values for features like virtual cards and child accounts) that are currently not offered within the app (see Table~\ref{tab:gap_analysis}). Equally important, the analysis reveals that some existing product areas are underrepresented in our intent taxonomy. For example, user queries signal a need for existing services such as managing language settings, managing payment methods, and tracking card status.

This targeted insight enables us to not only support the development of new offerings but also to refine and expand our search engine's service set. Integrating these data-driven insights ensures that our recommendations closely align with evolving user needs and expectations.

\section{Acknowledgments}
We would like to thank our linguist, James Garde, for his significant contributions on maintaining Chat146 / Chat186 and valuable feedback on our Query Data Enhancement system. We would also like to thank the Self Service Enablement team of JPMorganChase for supporting this research. 

\textbf{Disclaimer}: The research reported in this paper reflects the independent work and opinions of the authors. It is not representative of the views or official policies of JPMorganChase.

\bibliography{aaai2026}


\clearpage
\onecolumn

\appendix
\setcounter{figure}{0}
\renewcommand{\thefigure}{S\arabic{figure}}

\section{Synthetic Data Generation}

\subsection{Metrics for intrinsic synthetic data evaluation}
\label{intrinsic_sd_eval}

\subsubsection{Query length}
Query length refers to the average length of user queries, measured in characters. A longer sequence length might indicate more complex or detailed text, while a shorter sequence length could suggest more concise or simplified text.

\subsubsection{Distinct-\emph{N}}
The Distinct-\emph{N} score is calculated as the ratio of unique n-grams to the total number of n-grams in the dataset \citep{li_diversity-promoting_2016, li_contrastive_2023}. This score is averaged over different values of $n$ (i.e. from 1 to 4) to provide a comprehensive measure of diversity. It is defined as:
\begin{eqnarray}
\text{Distinct-\emph{N}} = \frac{1}{N} \sum_{n=1}^{N} \frac{\text{Number of Unique N-Grams}_n}{\text{Total Number of N-Grams}_n}
\end{eqnarray}
where $N$ is the maximum n-gram size considered.

\subsubsection{Self-BLEU}
Self-BLEU measures the diversity among the generated queries by treating each generated query as a candidate and using the remaining queries as references \citep{zhu_texygen_2018}. A high Self-BLEU score indicates that most generated queries share similar n-gram distributions, resulting in lower diversity. Conversely, a low Self-BLEU score suggests that the queries exhibit greater variability in wording and structure. The Self-BLEU score is computed as:
\begin{eqnarray}
\text{Self-BLEU} = \frac{1}{N} \sum_{i=1}^{N} \text{BLEU}(s_i, S \setminus \{s_i\})
\end{eqnarray}
where $N$ is the total number of generated queries, $s_i$ is the $i$th query, and $S \setminus \{s_i\}$ represents the set of all other generated queries used as reference.

\subsubsection{Query mean specificity}
Query mean specificity (QMS) is defined as the average inverse document frequency (IDF) of all terms in the corpus \citep{hutchison_empirical_2010}. It measures how unique or rare terms are within the corpus, with higher IDF values indicating terms that are more specific to the corpus and less common across documents. The IDF of a term $t$ is calculated as:
\begin{eqnarray}
\text{IDF}(t) = \log\left(\frac{\text{Number of Documents}}{\text{Number of Documents Containing } t}\right)
\end{eqnarray}

The average IDF score, representing the overall specificity, is calculated using the below formula where $T$ is the total number of terms:
\begin{eqnarray}
\text{QMS} = \frac{1}{T} \sum_{t \in \text{Terms}} \text{IDF}(t)
\end{eqnarray}

\subsubsection{Compression ratio}
The compression ratio (CR) is defined as the size of the compressed text divided by the size of the original text \citep{shaib_detection_2024}. This metric is computed using the compression algorithm gzip. A higher compression ratio suggests that the text is harder to compress, indicating greater complexity. It is given by:
\begin{eqnarray}
\text{CR} = \frac{\text{Size of Compressed Text}}{\text{Size of Original Text}}
\end{eqnarray}

\subsubsection{Compression ratio - part of speech}
The compression ratio - part of speech (CR-POS) is similar to the compression ratio but is applied to the sequence of POS tags derived from the text \citep{shaib_detection_2024}. This metric helps evaluate the diversity of sentence structures generated by the model. It is calculated as:
\begin{eqnarray}
\text{CR-POS} = \frac{\text{Size of Compressed POS Tags}}{\text{Size of Original POS Tags}}
\end{eqnarray}

\subsection{Additional Experiments}

\begin{figure}[H]
\centering
\includegraphics[width=0.8\textwidth]{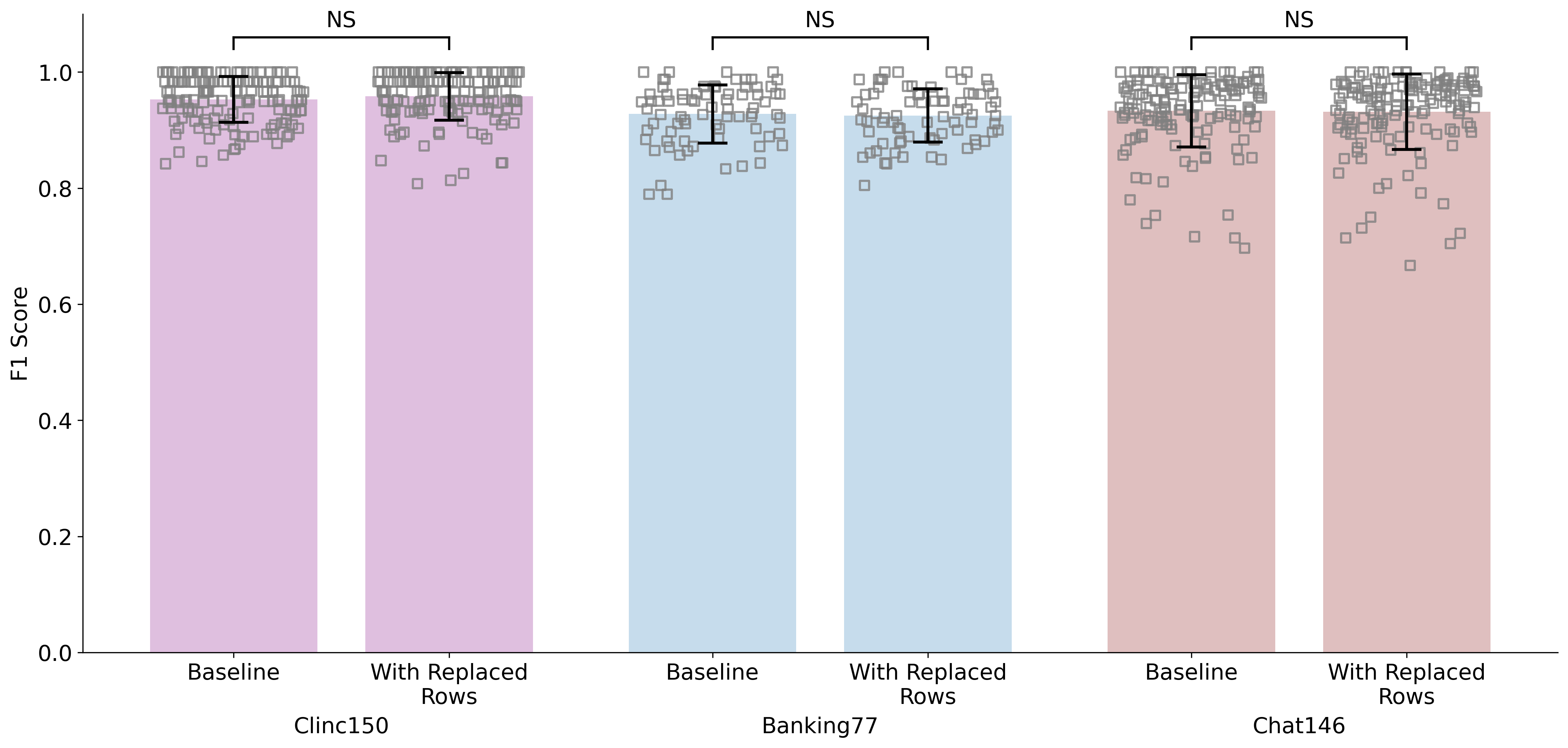}
\caption{This figure compares the macro-averaged F1 scores (mean $\pm$ std) between models trained on the original (control) data and those in which 25\% of the training queries were replaced with synthetic queries. The results demonstrate that simply substituting a portion of real queries with synthetic ones does not significantly affect classifier performance across all three datasets, thereby rationalizing our focus on the more challenging "cold start" scenario of complete label replacement.}
\label{fig:generation_results_rows}
\end{figure}
\label{synth_row_replacement}

\begin{figure}[H]
\centering
\includegraphics[height=7cm]{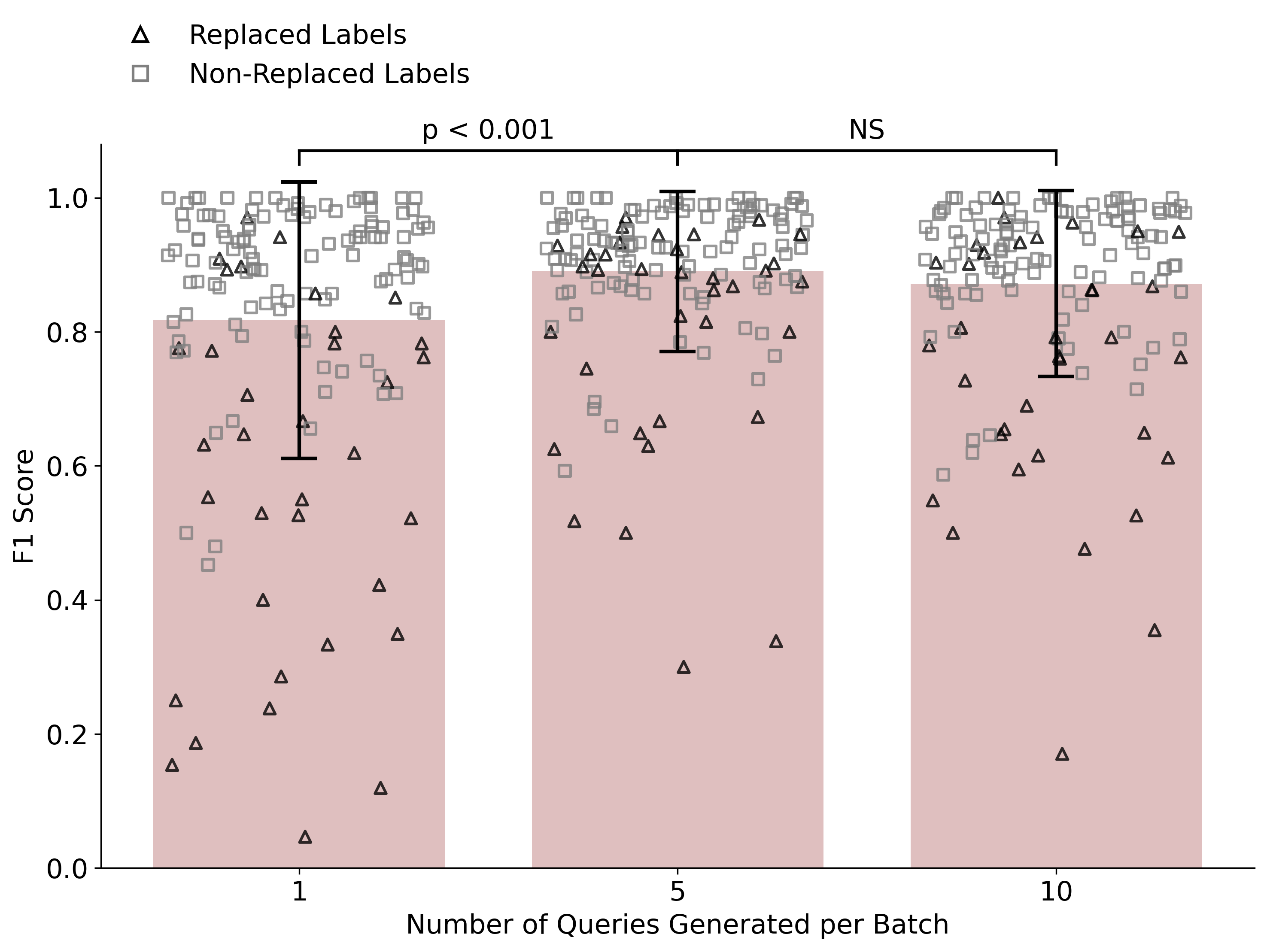}
\caption{Chat146 – query batch experiment. This figure illustrates the impact of different query batch counts on the performance of Chat146 models under the condition where synthetic queries completely replaced all training queries for 25\% of labels (mean $\pm$ std). The results demonstrate that our sequential conditioning approach yields a significant performance benefit: generating 100 queries in batches of 5 (default) leads to better model performance compared to generating 100 individual queries.}
\label{fig:generation_results_batch}
\end{figure}
\label{synth_batch}

\begin{figure}[H]
\centering
\includegraphics[height=7cm]{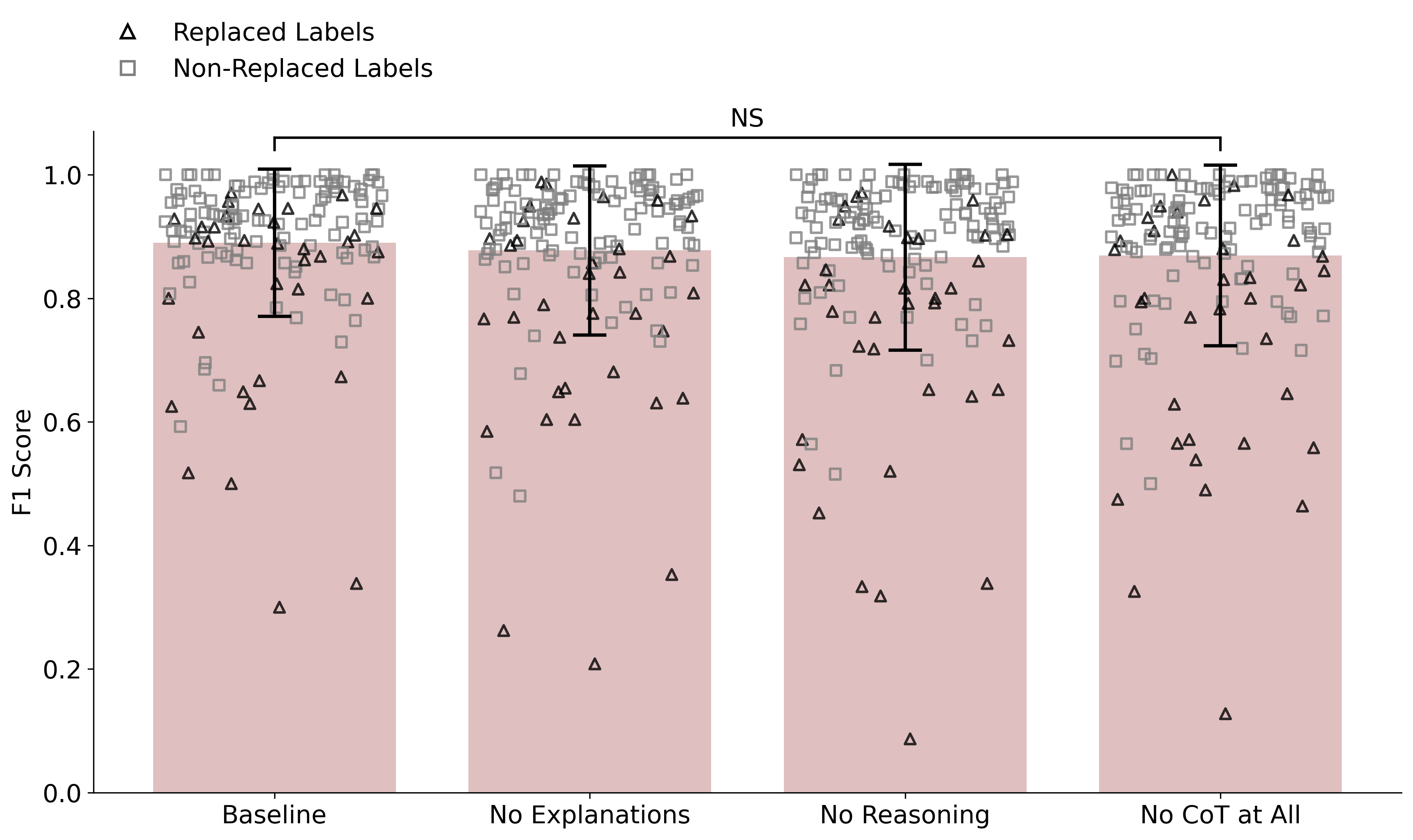}
\caption{Chat146 – chain of thought experiment. This figure shows the effect of omitting various chain-of-thought elements from the Synthetic Query Generation prompt on the performance of Chat146 models under the condition where synthetic queries completely replaced all training queries for 25\% of labels (mean $\pm$ std). While the removal of these structured output elements did not result in statistically significant changes in macro-averaged F1 scores, the highest performance was achieved when the chain-of-thought elements were included (default).}
\label{fig:generation_results_cot}
\end{figure}
\label{synth_cot}

\begin{figure}[H]
\centering
\includegraphics[height=7cm]{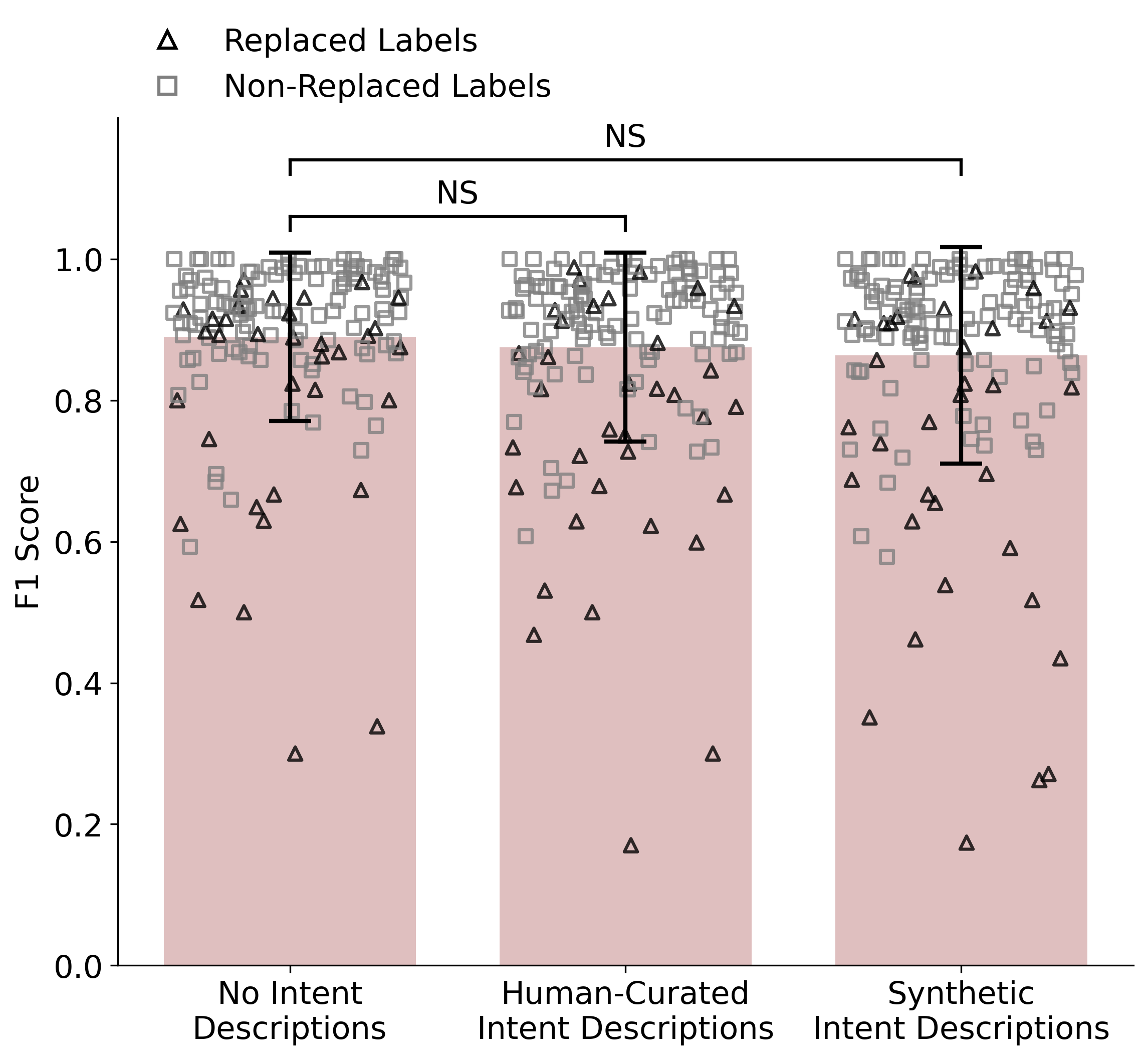}
\caption{Chat146 – prompt description experiment. This figure shows the impact of three synthetic query prompt strategies on the performance of Chat146 models under the condition where synthetic queries completely replaced all training queries for 25\% of labels (mean $\pm$ std). The prompt strategies are to include: (i) no description in the prompt (default), (ii) a human-curated description, and (iii) an LLM-generated description derived from the 10 few-shot query examples and also 10 few-shot description examples. The findings indicate that the inclusion or type of description does not produce a statistically significant change in classifier performance.}
\label{fig:generation_results_desc}
\end{figure}
\label{synth_desc_results}

\begin{figure}[H]
\centering
\includegraphics[height=7cm]{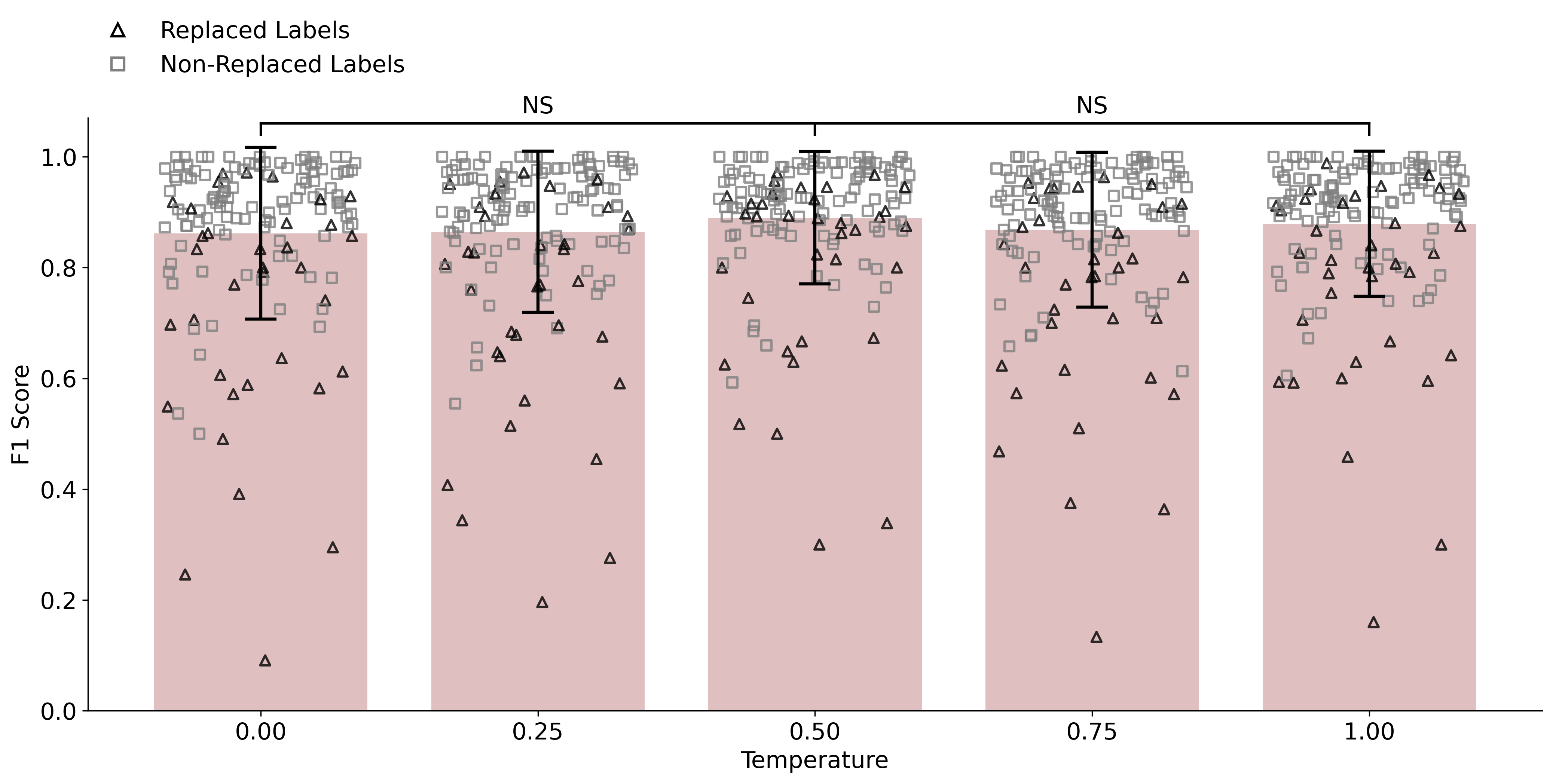}
\caption{Chat146 – temperature experiment. This figure illustrates the impact of different temperature settings on the performance of Chat146 models under the condition where synthetic queries completely replaced all training queries for 25\% of labels (mean $\pm$ std). Although varying the temperature did not yield statistically significant changes in macro-averaged F1 scores, the highest performance was achieved at a temperature of 0.5 (default).}
\label{fig:generation_results_temp}
\end{figure}
\label{synth_temp_results}

\begin{figure}[H]
\centering
\includegraphics[height=7cm]{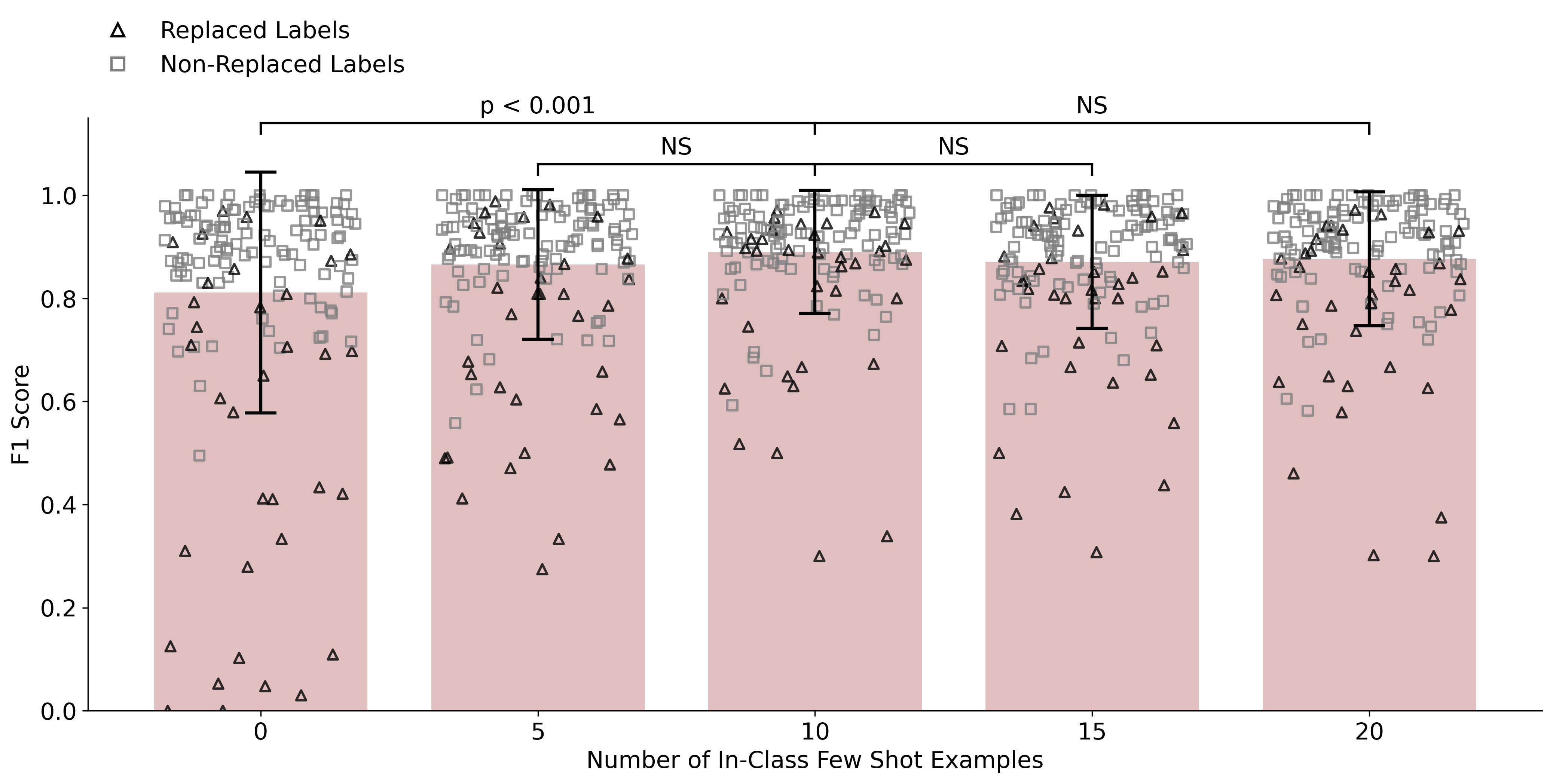}
\caption{Chat146 – few-shot examples experiment. This figure evaluates how the number of in-class few-shot examples influences the performance of Chat146 models under the condition where synthetic queries completely replaced all training queries for 25\% of labels (mean $\pm$ std). The macro-averaged F1 scores reveal that incorporating in-class few-shot examples significantly improves performance; however, the differences between using 5 versus 20 examples were not statistically significant (default was 10).}
\label{fig:generation_results_fs}
\end{figure}
\label{synth_fs_results}

\begin{figure}[H]
\centering
\includegraphics[height=7cm]{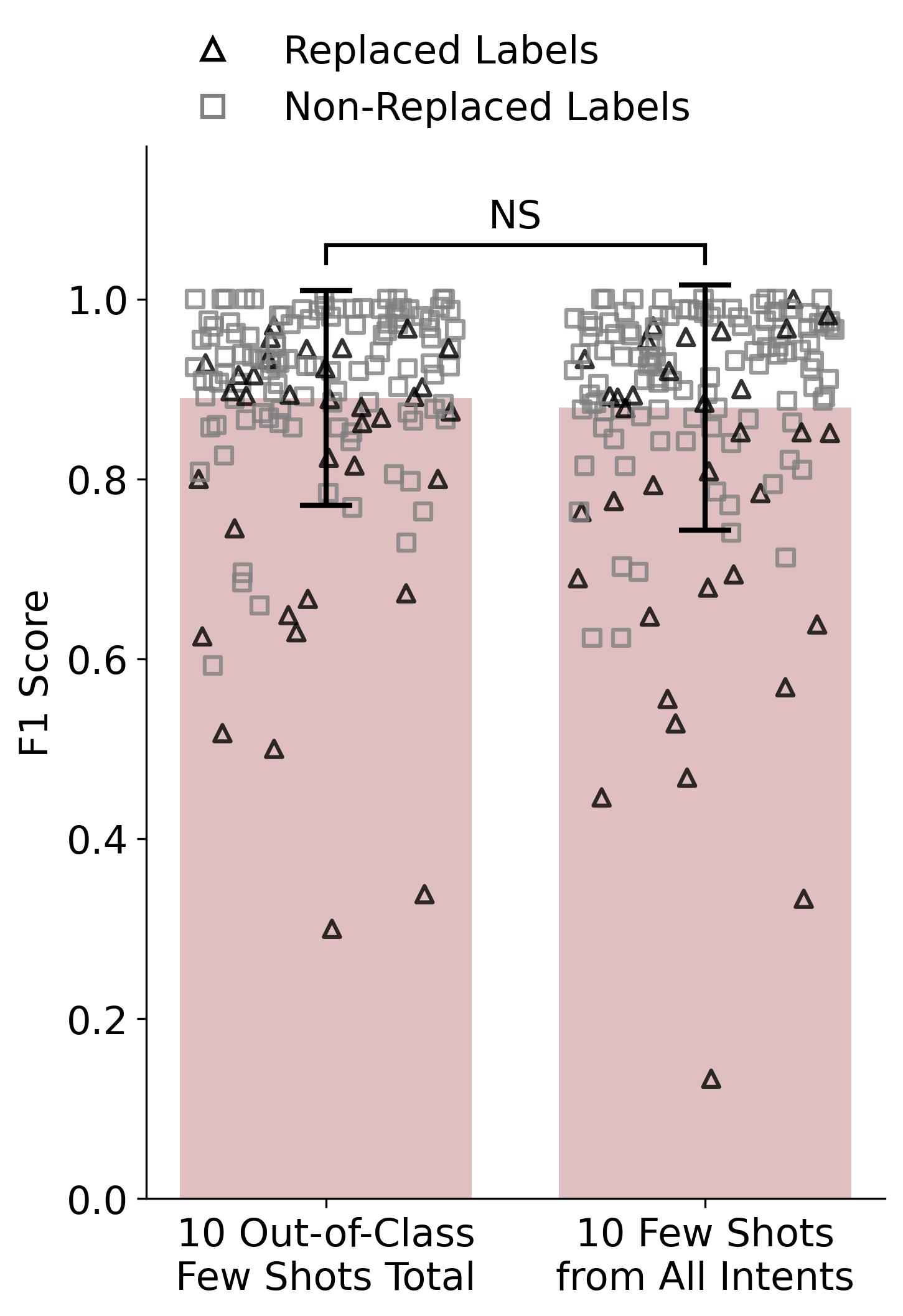}
\caption{Chat146 – intent context experiment. This figure shows the impact of two synthetic query prompt strategies on the performance of Chat146 models under the condition where synthetic queries completely replaced all training queries for 25\% of labels (mean $\pm$ std). The prompt strategies are: (i) include 10 randomly sampled few-shot examples from other intents (default), and (ii) include 10 few-shot examples from every intent in the taxonomy. The results indicate that providing examples from the entire set of intents does not offer a statistically significant performance benefit over the approach using randomly sampled examples. This ultimately suggests that simply including examples from all other intents as context isn’t enough to prevent overlap between synthetic queries and similar existing intents, which was observed to negatively impact performance.}
\label{fig:generation_results_tax}
\end{figure}
\label{synth_fc_results}

\subsection{Prompts}
\label{synth_prompt}
\begin{tcolorbox}[breakable, colback=gray!10, colframe=black, title=Synthetic Data Generator]
You are an expert banking search assistant query generator for JPMorgan Chase.

Your task is to create realistic user queries that customers might type when navigating to specific features in the banking app.

IMPORTANT GUIDELINES:

1. Create diverse, natural-sounding customer queries.

2. Include queries that vary substantially in length --- from very short fragments and abbreviations (common in real mobile text inputs) to slightly longer, detailed questions.

3. Vary query complexity, specificity, and length.

4. Consider both direct commands and conversational, informal questions.

5. Focus on the core functionality users are trying to access with each query.

6. Be as concise and specific as the given examples, while also presenting a realistic diversity in users' query style.

7. Some intents capture vague or one-word queries; others are more specific.

8. Queries should mimic the casual and conversational tone of real user queries.

You will be given a user intent or navigation key (navkey) from the JPMorgan Chase banking app.

Your task is to generate 5 diverse user queries that would naturally lead to this navkey/intent.

These queries should capture the natural variability observed in real interactions:

- Varying query lengths (from one or two words to full sentences).

- Realistic abbreviations or typographical errors.

- Some queries may be ambiguous or lack details, reflecting genuine user uncertainty.

- Others should be more detailed when necessary.

Context \& Examples:

Full Taxonomy Examples (if provided):

\texttt{\{full\_taxonomy\}}

Intent:  

\texttt{\{intent\}}

Description (if provided):  

\texttt{\{description\}}

Few-shot Examples (if provided):

\texttt{\{few\_shot\_examples\}}

Before generation, reflect on the observed characteristics:

- Real queries are notably short, often with missing sentence structures.

- There is continuous variation between imperatives and questions.

- Include natural misspellings and incomplete phrases to simulate mobile input.

Then proceed to generate.

Make sure the queries:

- Vary in length, from short fragments to full sentences.

- Include both questions and commands.

- Use natural, conversational language with occasional typographical errors.

- Reflect how real banking customers phrase their requests.

Return 5 user queries in YAML format (respond only in YAML):

label: "$<$Provided\_Intent\_Label$>$"

reflection: "$<$Step-by-step reflection on what the intent captures regarding customer needs and query generation diversity$>$"

generated\_queries:

  - reasoning: "$<$Step-by-step planning for generating the first query$>$"
  
    text: "$<$First user query generated based on reasoning$>$"
    
    explanation: "$<$Detailed explanation why this generated query is appropriate and distinct$>$"

  - reasoning: "$<$Step-by-step planning for generating the second query$>$"
  
    text: "$<$Second user query generated based on reasoning$>$"
    
    explanation: "$<$Detailed explanation why this generated query is appropriate and distinct$>$"

  - reasoning: "$<$Step-by-step planning for generating the third query$>$"
  
    text: "$<$Third user query generated based on reasoning$>$"
    
    explanation: "$<$Detailed explanation why this generated query is appropriate and distinct$>$"

  - reasoning: "$<$Step-by-step planning for generating the fourth query$>$"
  
    text: "$<$Fourth user query generated based on reasoning$>$"
    
    explanation: "$<$Detailed explanation why this generated query is appropriate and distinct$>$"

  - reasoning: "$<$Step-by-step planning for generating the fifth query$>$"
  
    text: "$<$Fifth user query generated based on reasoning$>$"
    
    explanation: "$<$Detailed explanation why this generated query is appropriate and distinct$>$"

\end{tcolorbox}

\section{Intent Disambiguation}

\subsection{Prompts}
\label{id_prompt}

\begin{tcolorbox}[breakable, colback=gray!10, colframe=black, title=Data Labeler Prompt]

You are an expert data labeler at JPMorgan Chase

\texttt{subintent\_context}

I have given you above a taxonomy of labels and example user queries for that label.

Label just the below \texttt{num\_queries} user queries based on the taxonomy.

For context, the default class of this data is \texttt{default\_label} (i.e. the utterance does not match any of the other labels or is vague).

\texttt{user\_queries\_str}

Respond with just a YAML list of queries (exactly as given) and their labels.

First reason about the given examples and how they relate to the examples in the taxonomy.

If you cannot confidently label a query, label it with the default class of this data.

The responses must be in order — i.e. the first response corresponds to the first input query, and so on.

Example YAML format:

reasoning: "$<$Quick reasoning about the given queries and taxonomy$>$"

queries:

- text: "$<$text\_1$>$"

  label: "$<$label\_1$>$"

- text: "$<$text\_2$>$"

  label: "$<$label\_2$>$"

...

\end{tcolorbox}

\section{Intent Gap Analysis}

\subsection{Additional Experiments}
\label{IGA_merge_figs}

\begin{figure}[H]
\centering
\includegraphics[width=0.95\textwidth]{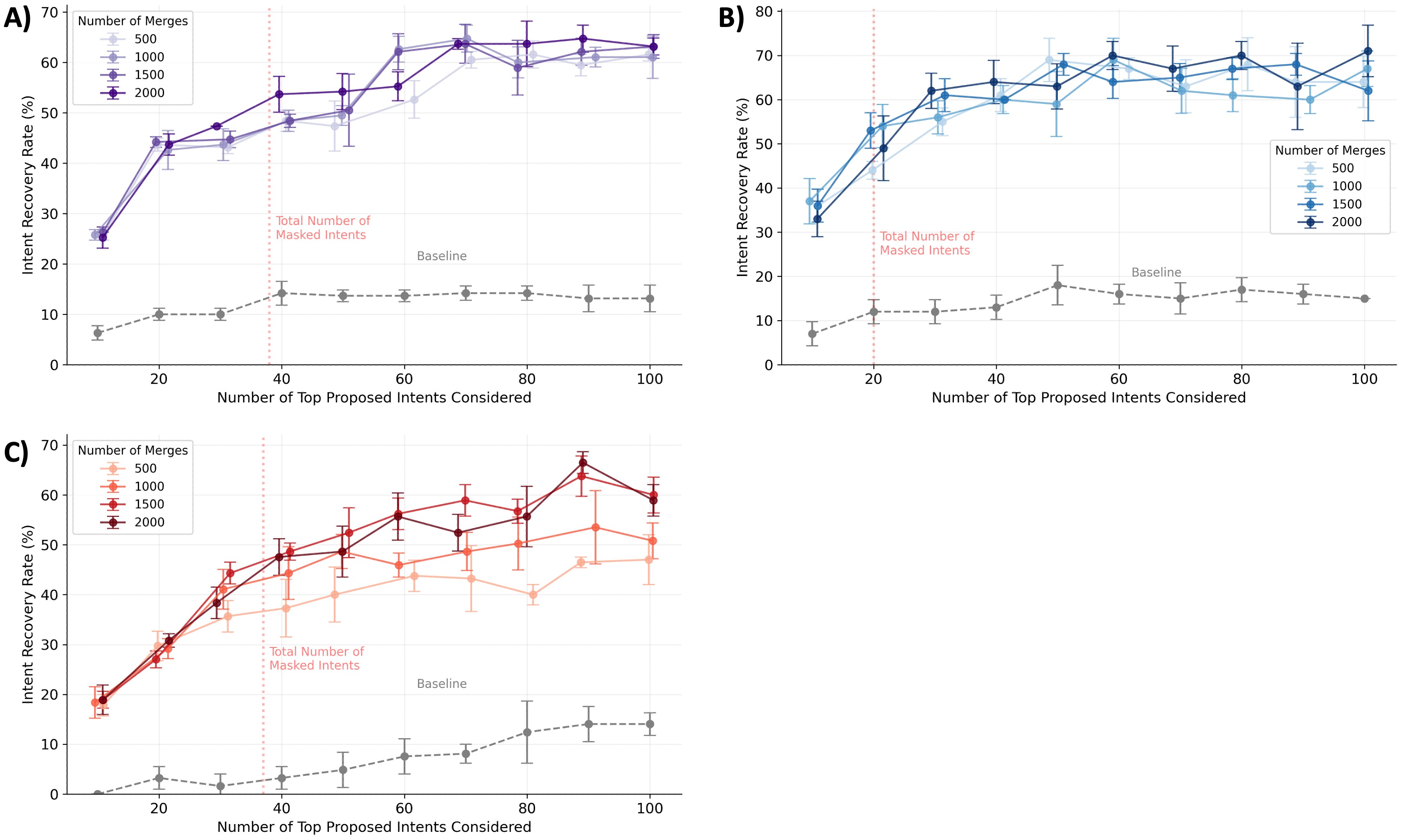}
\caption{Rediscovery of masked intents results. After masking 25\% of the intents (with their few-shot examples) from the known set, we applied our method to evaluate how effectively the masked intents can be rediscovered. This figure showcases the effect of the number of merges attempted by the Intent Merger agent on intent recovery rate. Panel A corresponds to the Clinc150 dataset, panel B to Banking77, and panel C to Chat146. This figure shows that the number of merges does not significantly improve the intent recovery rate past 1000 attempts. However, continuing merges past this point builds additional support for the most frequently proposed intents, which is helpful from a business context.}
\label{fig:rediscovery_results_all}
\end{figure}

\subsection{Prompts}
\label{iga_prompts}
\begin{tcolorbox}[breakable, colback=gray!10, colframe=black, title=Intent Proposer]
You are tasked with identifying gaps in our existing customer intent taxonomy using real user queries.

For each query, determine if it clearly falls outside our current categories by considering its linguistic nuances and the practical needs for dedicated digital solutions (e.g., dedicated pages or search options).

Existing user intents with examples:

\texttt{taxonomy\_df\_str}

Real user queries:

\texttt{queries\_subset\_str}

Your task is to perform a gap analysis comparing the provided real user queries against the existing customer intent taxonomy.

Evaluate each query through both a linguistic lens and a practical website design perspective to determine if it signifies a genuine gap—indicating an unaddressed customer need.

Instructions:

1. Analyze each user query and assess whether it clearly fits into one of the existing intent categories. Focus on nuanced language, context, and semantic meaning.

2. Identify queries that, despite potential similarities, demand a separate, actionable intent that would justify the creation of dedicated webpages or search options.

3. For every gap identified, provide:

  - The exact example query (as it appears in the provided data).
  
  - A proposal for a new intent capturing the emerging customer need.
  
4. The chain-of-thought reasoning must include your logical steps: explain how you determined that the query is distinct from current intents based on linguistic nuances and its potential impact on digital design.

5. Note that "inquiry" or "general" intents are unacceptable; the proposed intents must be specific and actionable.

6. Given that the current taxonomy is fairly comprehensive, only include examples that clearly cannot be mapped to any existing categories.

7. If none of the queries meet these criteria, set Valid to False.

Please provide your response by filling out the following YAML template (respond only in valid YAML):

Reasoning: $<$A detailed analysis and explanation for why the selected examples could not be classified under the current intents, including considerations of linguistic nuances and digital design requirements$>$

Examples:

  - example: $<$A specific customer utterance from the given examples that is not covered by existing intents$>$
  
    proposed\_intent: $<$New intent proposal capturing the customer need$>$
    
  - example: $<$Another valid gap example$>$
    proposed\_intent: $<$New intent proposal capturing the customer need$>$
    
...

Valid: $<$True if at least one genuine gap is identified based on the criteria, otherwise False$>$

\end{tcolorbox}

\begin{tcolorbox}[breakable, colback=gray!10, colframe=black, title=Intent Judge]
You are tasked with evaluating a proposed customer intent against our existing customer intent taxonomy.

Your goal is to classify a single proposed intent into one of four categories based on its novelty, relevance, and potential impact on our digital offerings.

The current customer intent taxonomy with examples:

\texttt{taxonomy\_df\_str}

Proposed intent:

\texttt{proposed\_intent\_str}

Example query or queries for the proposed intent:

\texttt{example\_queries\_str}

Your task is to evaluate the proposed intent and classify it into one of the following categories:

1. Not novel / handled already by an existing intent (return "not\_novel")

2. Can be easily answered by an FAQ (return "faq")

3. Not relevant to banking/financial products and services (return "not\_relevant")

4. A product, service, or experience that we should consider adding to the Chase mobile app (return "consider\_adding")

Instructions:

1. Analyze the proposed intent and its example query (or queries) in the context of the existing taxonomy.

2. Determine if the proposed intent is already covered by an existing intent. If so, classify it as "not\_novel."

3. If the proposed intent can be addressed through a simple FAQ, classify it as "faq."

4. If the proposed intent represents a significant new customer need that warrants the development of a new product, service, or experience, classify it as "consider\_adding."

5. If the proposed intent is not relevant to banking or financial products and services, classify it as "not\_relevant."

6. Provide a brief explanation for your classification decision, considering linguistic nuances, customer needs, and potential digital design implications.

7. Mention the top most similar intents from the existing taxonomy in your reasoning.

Please provide your response by filling out the following YAML template (respond only in valid YAML):

Reasoning: $<$Think about the proposed intent and your classification decision, including the top most similar intents$>$

Classification: $<$One of the categories: "not\_novel", "faq", "not\_relevant", "consider\_adding"$>$.

\end{tcolorbox}

\begin{tcolorbox}[breakable, colback=gray!10, colframe=black, title=Intent Merger]

\texttt{\{intent\_mapping\_str\}}

I generated a set of proposed intents from customer queries. Each entry above is an intent followed by its examples.

Your task is to analyze these intents and identify a single pair that is semantically redundant (if present).  

That is, two intents with examples that are indistinguishable and would result in the creation of the same webpage or search option.

Please adhere to the following rules:

1. Identify pairs of intents whose examples are essentially identical.

2. Only merge intents when:

   - The language and examples are exactly the same.
   
   - The end-user experience, in terms of webpage content or search functionality, would be identical.
   
   - If there is any nuance suggesting that two intents serve different purposes, keep them separate.

3. If no single pair of intents meets these strict criteria, return "False" for the "Valid" field.

Return your analysis in the exact YAML format below (do not include any extra text):

Reasoning: $<$Detailed reasoning on why the chosen intents are or are not redundant based on their examples and linguistic nuances$>$

Pair: $<$(intent1, intent2)$>$   \# The pair of intents that are semantically redundant

Keep: $<$intent to keep$>$           \# Indicate which intent should be retained

Keep Examples:

  - $<$example 1 exactly as provided$>$
  
  - $<$example 2 exactly as provided$>$
  \dots

Eliminate: $<$intent to eliminate$>$ \# Indicate the redundant intent to remove

Eliminate Examples:

  - $<$example 1 exactly as provided$>$
  
  - $<$example 2 exactly as provided$>$
  \dots

Valid: $<$True/False$>$           \# True if a merge is justified based on the criteria; otherwise False

\end{tcolorbox}

\begin{tcolorbox}[breakable, colback=gray!10, colframe=black, title=Masked Intent Rediscovery Evaluator]
You are tasked with determining whether a single given intent is already present in a set of proposed intents.

Below, you'll find additional context for the intent including its description and some example queries from the original labeled data.

Please analyze whether the semantic meaning, language, and context of the intent is also captured and represented in at least one of the proposed intents.
    
Proposed Intents:  
\texttt{\{proposed\_intents\_str\}}

Intent:  
\texttt{\{masked\_intent\}}

Description:  
\texttt{\{description\}}

Few-shot Examples:  
\texttt{\{examples\_text\}}

Instructions:

1. Analyze the dropped intent along with its description and examples.

2. Compare its core meaning with each of the proposed intents and decide if it is effectively covered by one of the proposed intents.

3. Provide a detailed reasoning about any semantic similarities or differences.

Return your result using the following YAML format (respond only in YAML):

Reasoning: $<$Detailed explanation including the similarities or differences between the intent's context and the proposed intents$>$

Given: $<$Just the original intent, do not include its description$>$

Most Similar: $<$The most similar proposed intent to the original intent$>$

Match: $<$True/False; True if the given intent is covered by one of the proposed intents, otherwise False$>$

\end{tcolorbox}

\end{document}